\documentclass[%
reprint,
superscriptaddress,
 amsmath,amssymb,
 aps,
pra,
]{revtex4-2}
\usepackage{graphicx}
\usepackage{dcolumn}
\usepackage{bm}
\usepackage{amsmath}
\usepackage{amssymb}
\usepackage{braket}
\usepackage{mathrsfs}
\usepackage{upgreek}

\begin{document}
\preprint{APS/123-QED}

\title{Photoswitchable radicals as reporter spins for quantum sensing with spin defects in diamond}

\author{Lakshmy Priya Ajayakumar}
 \affiliation{Department of Chemistry, University of Illinois at Urbana-Champaign, Urbana, Illinois, USA}
 \affiliation{Illinois Quantum Information Science and Technology Center (IQUIST), University of Illinois at Urbana-Champaign, Urbana, Illinois, USA}

\author{David J. Durden}
\affiliation{Department of Chemistry, University of Illinois at Urbana-Champaign, Urbana, Illinois, USA}
\affiliation{Illinois Quantum Information Science and Technology Center (IQUIST), University of Illinois at Urbana-Champaign, Urbana, Illinois, USA}

\author{Aksshay Nandakumar Regeni}
\affiliation{Department of Chemistry, University of Illinois at Urbana-Champaign, Urbana, Illinois, USA}
\affiliation{Illinois Quantum Information Science and Technology Center (IQUIST), University of Illinois at Urbana-Champaign, Urbana, Illinois, USA}

\author{Mingcai Xie}
\affiliation{Department of Chemistry, University of Illinois at Urbana-Champaign, Urbana, Illinois, USA}
\affiliation{Illinois Quantum Information Science and Technology Center (IQUIST), University of Illinois at Urbana-Champaign, Urbana, Illinois, USA}

\author{Swastik Hegde}
\affiliation{Center for Biophysics, University of Illinois at Urbana-Champaign, Urbana, Illinois, USA}
\affiliation{Illinois Quantum Information Science and Technology Center (IQUIST), University of Illinois at Urbana-Champaign, Urbana, Illinois, USA}

\author{Gustavo Aldas}
\affiliation{Department of Chemistry, University of Illinois at Urbana-Champaign, Urbana, Illinois, USA}
\affiliation{Illinois Quantum Information Science and Technology Center (IQUIST), University of Illinois at Urbana-Champaign, Urbana, Illinois, USA}

 \author{Kyle Haggard}
\affiliation{Department of Chemistry, University of Illinois at Urbana-Champaign, Urbana, Illinois, USA}
\affiliation{Illinois Quantum Information Science and Technology Center (IQUIST), University of Illinois at Urbana-Champaign, Urbana, Illinois, USA}

\author{Mikael P. Backlund}%
 \email{mikaelb@illinois.edu}
 \affiliation{Department of Chemistry, University of Illinois at Urbana-Champaign, Urbana, Illinois, USA}
\affiliation{Illinois Quantum Information Science and Technology Center (IQUIST), University of Illinois at Urbana-Champaign, Urbana, Illinois, USA}
\affiliation{Center for Biophysics, University of Illinois at Urbana-Champaign, Urbana, Illinois, USA}





\date{\today}

\begin{abstract}
The rapid decay of target signal strength with distance from the sensor presents a key challenge in nanoscale magnetic sensing with  nitrogen-vacancy (NV) centers in diamond, limiting both sensitivity and spatial resolution. Here we introduce a strategy to overcome this limitation by using radical anions formed from rhodamine-derived dyes as reporter spins localized to the diamond surface. These radicals, generated through photoreduction, are optically identifiable and stable on timescales exceeding an hour. We experimentally demonstrate their coherent manipulation and detection using single, shallow NV centers for readout. We observe heterogeneity in the local magnetic environments of the photoactivated spins from site to site, likely due to variations in inter-radical couplings across our measurements. Looking forward, our approach enables correlative nanoscale magnetic and optical imaging, and opens new pathways toward single-molecule magnetic resonance studies.  


\end{abstract}

\maketitle


\section{\label{sec:level1}INTRODUCTION}

Advances in quantum control of nitrogen vacancy (NV) centers in diamond have driven remarkable progress in nanoscale spin detection and magnetic resonance spectroscopy. With their optical addressability \cite{jelezko_observation_2004, schirhagl_nitrogen-vacancy_2014, heremans_control_2016}, high sensitivity  \cite{taylor_high-sensitivity_2008, muller_nuclear_2014, zopes_three-dimensional_2018}, nanoscale spatial resolution \cite{maze_nanoscale_2008, grinolds_subnanometre_2014,devience_nanoscale_2015,abobeih_atomic-scale_2019}, and operation under ambient conditions \cite{taylor_high-sensitivity_2008,maze_nanoscale_2008, balasubramanian_nanoscale_2008}, NV centers have enabled nuclear magnetic resonance (NMR) spectroscopy of single proteins \cite{lovchinsky_nuclear_2016} as well as electron paramagnetic resonance (EPR) detection of individual spin labels affixed to biomolecules \cite{sushkov_all-optical_2014, shi_single-dna_2018}. In condensed matter systems, advanced NV sensing protocols have unlocked new possibilities in imaging magnetic domain walls in thin ferromagnets \cite{tetienne_nanoscale_2014}, mapping vortex cores in superconductors \cite{schlussel_wide-field_2018}, and probing spin-wave excitations in two-dimensional quantum magnets \cite{bertelli_magnetic_2020}. Due to the characteristic falloff of magnetic dipole-dipole interactions, achieving nanoscale resolution and single-spin sensitivity in these applications critically depends on minimizing the NV-sample distance \cite{degen_nanoscale_2009, staudacher_nuclear_2013, mamin_nanoscale_2013}. Consequently, NV centers must often be placed no more than a few nanometers beneath the diamond surface. At such shallow depths, however, charge instability and surface-induced decoherence begin to outweigh the advantages of proximity to the source \cite{myers_probing_2014, sangtawesin_origins_2019, rondin_surface-induced_2010,  yuan_charge_2020, rosskopf_investigation_2014}. Overcoming these surface-related constraints remains a central challenge for advancing NV-based nanoscale magnetometry beyond proofs-of-principle, and in turn realizing its full potential across physics, chemistry, and biology. 

A promising strategy to address this challenge is the use of external electronic “reporter” spins that act as intermediaries between shallow NV centers and external targets \cite{sushkov_magnetic_2014, zhang_reporter-spin-assisted_2023}. With magnetic moments orders of magnitude larger than those of nuclear spins, such reporter spins can enhance NV-target coupling while remaining coherently controllable and readable via dipolar interactions with the NV \cite{grotz_sensing_2011, mamin_detecting_2012}. Theoretical studies have suggested that such architectures could enable single-nuclear-spin detection and imaging, with subnanometer spatial resolution by spin amplification \cite{schaffry_proposed_2011}. A previous implementation demonstrated the feasibility of reporter-assisted sensing employing surface-bound electronic spin impurities to localize surface-bound proton impurities \cite{sushkov_magnetic_2014}. The generality of this approach is hindered by the instability \cite{dwyer_probing_2022} and lack of optical control of these ``dark spins'', however, motivating the search for an alternative platform.

\begin{figure*}[t]
\centering
\includegraphics[width=15.5cm,height=11.4cm]{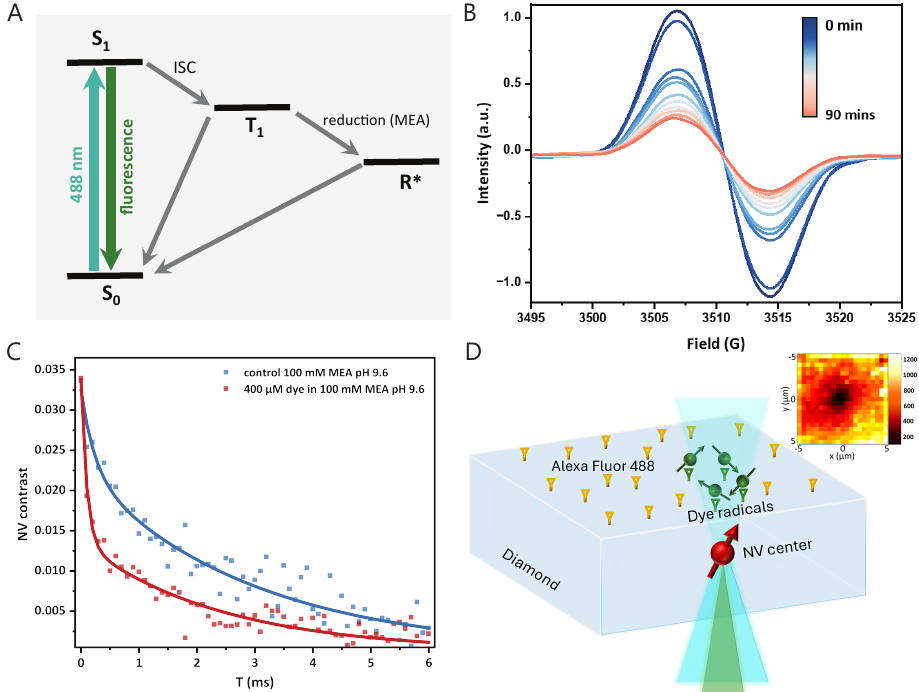}
\caption{Dye-derived radicals as reporter spins. (A) Energy-level diagram of Alexa Fluor 488 depicting photoreduction of dye triplet state to form radical anions. (B) Room-temperature cw-EPR spectra of an aqueous solution of 400-$\upmu$M Alexa Fluor in 100 mM MEA at pH 9.6 after irradiation with 488-nm light for 5 minutes. (C) Relaxometry measurements using a dense, shallow ensemble of NVs, with (red) and without (blue) photoactivated dye present in the solution above. Solid lines are bi-exponential fits to the data. (D) Illustration of Alexa Fluor 488 molecules attached to the diamond surface forming radicals upon 488-nm photoexcitation in the presence of MEA. The inset shows a confocal image of the dye-functionalized diamond surface after burning a hole with the laser (fluorescence units arbitrary).}
\label{fig:Fig1}
\end{figure*}

Here, we propose the use of long-lived organic radicals derived from fluorescent dye molecules as independently addressable, photoswitchable reporter spins. Rhodamine and its derivatives are known to form radical anions under certain conditions \cite{beaumont_excited_1997,zondervan_photobleaching_2004}. In the presence of a mild reducing agent, popular rhodamine-based fluorophores including many Alexa and ATTO dyes can form stable (lasting up to hours) radical anions upon quenching of their fluorescence \cite{van_de_linde_photoinduced_2011}. In other words, coincident with a turn-off in fluorescence, these molecules acquire an electronic spin of $S=1/2$. Moreover, the process is reversible and repeatable, with return to the fluorescent state facilitated by either exposure to oxygen or illumination with 405-nm light \cite{van_de_linde_photoinduced_2011}. The mechanism of photoreduction is shown in Fig. \ref{fig:Fig1}A. The reaction is initiated by an intersystem crossing (ISC) from the excited singlet state to the triplet manifold. In the presence of mercaptoethylamine (MEA) the triplet state is quenched, leading to the formation of stable radical anions in room-temperature aqueous solution. The electron donor must be present in excess compared to dissolved molecular oxygen [$\sim200$ $\upmu$M \cite{gorner_oxygen_2008}] in order to outpace the transition from the triplet directly back to the singlet ground state. We note that MEA and similar mild reducing agents are routinely employed in single-molecule and super-resolution fluorescence microscopy as a key component of so-called reducing and oxidizing systems (ROXS), used to tune photostability and blinking \cite{vogelsang_reducing_2008,heilemann2008subdiffraction,heilemann2009super,zheng2014ultra}.

The prospect of leveraging this behavior for nanoscale magnetometry offers several enticing capabilities. For one, the ability to independently localize the spin-procuring molecules via their optical emission could significantly speed up identification of suitable sensor-reporter pairs, obviating the need for lengthy trial-and-errors searches. Precise co-localization of the spin beacon and NV would provide useful prior information for interpreting magnetic resonance data, as magnetic dipole-dipole coupling is sensitive to both separation distance and the angle made between the separation vector and the direction of the applied field. The full toolbox of single-molecule fluorescence microscopy presents additional routes to such prior information, as conveyed, for instance, by the orientation and rotational mobility of a switchable dye affixed to a target biomolecule \cite{brasselet2025single}.

In this work we experimentally demonstrate the ability to coherently interface nanoscale ensembles (on the order of tens) of these spins with single NV sensors via double electron-electron resonance (DEER) spectroscopy. Our measurements reveal heterogeneity in the local magnetic environment of the reporter spins from site to site, likely resulting from stochastic variation in the positions and therefore inter-spin couplings of the surface-bound qubits. Our results suggest a new path forward for nanoscale and single-molecule magnetic resonance microscopy.

\section{RESULTS AND DISCUSSION}
\subsection{Ensemble measurements}
For our experiments, we selected Alexa Fluor 488, whose fluorescence spectrum does not significantly overlap with that of the NV center. Continuous-wave electron paramagnetic resonance (cw-EPR) measurements on a solution containing $\sim400$ $\upmu$M dye at pH 9.6 (unsealed, and at room temperature) confirmed the formation of these dye radicals upon 488 nm laser irradiation for 5 minutes (Fig. \ref{fig:Fig1}B). A series of bulk EPR recordings showed a significant fraction of these radicals survived for longer than an hour, long enough, in principle, to avail themselves to repeated interrogation by NV centers.

Next we performed relaxometry using a shallow, dense ensemble of NV centers, both in the presence and absence of activated dye. First, we deposited 100 mM MEA solution at pH 9.6 without dye on the diamond chip and measured the NV ensemble's $T_1$ relaxation curve (Fig. \ref{fig:Fig1}C, blue). The decay includes a fast (sub-ms) component, possibly due to interactions with a small fraction of rapidly relaxing charge fluctuations, as reported previously in very dense NV ensembles \cite{PhysRevLett.118.093601}. Next, we displaced the solution with one also containing 400 µM Alexa 488. We irradiated continuously with a 488-nm laser for 5 minutes, then performed another measurement of the NVs' $T_1$ relaxation (Fig. \ref{fig:Fig1}C, red). Magnetic noise due to the formed radicals clearly accelerates the NV relaxation, despite the presence of the competing fast-relaxation component.

\subsection{Single-NV DEER experiments}
For most of our subsequent experiments targeting single NVs, the surface of a diamond chip containing individually resolvable NV centers with a mean implantation depth around 12 nm was functionalized with Alexa Fluor 488 dye molecules (Fig. \ref{fig:Fig1}D) via carbodiimide cross-coupling (see Materials and Methods for details). The functionalized diamond chip was placed in a channel slide and 50-mM MEA solution at pH 8 was flowed in. We avoided higher pH in our single-NV experiments in order to minimize pH-induced charge instability of the NVs \cite{sow_high-throughput_2020}. Radical formation on the diamond surface was induced by irradiation with a 488 nm laser for a duration of $\sim30-120$ seconds. The inset of Fig. \ref{fig:Fig1}D depicts a hole burned into a sheet of fluorescence resulting from such an irradiation period. Fluorescence filters and laser lines were then switched out as a single NV coinciding with the 488-exposed region was identified and targeted.

\begin{figure}[t]
\centering
\includegraphics[width=8.5cm,height=10.389cm]{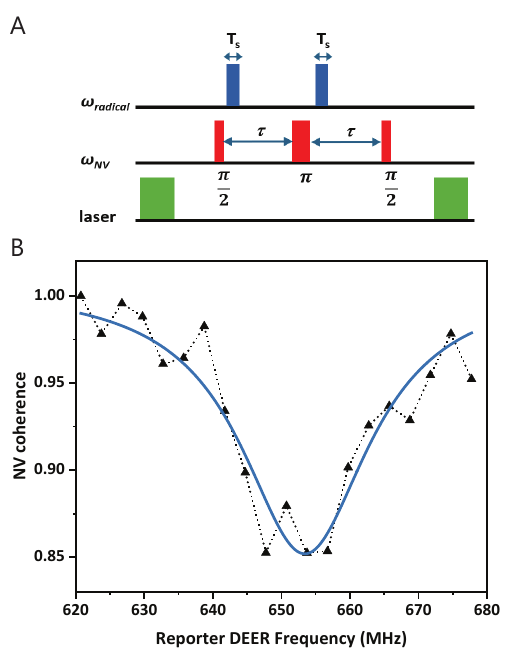}
\caption{Sensing radical reporter spins using a single, shallow NV center. (A) DEER pulse sequence. Two pulses of duration $T_s$ and frequency $\omega_{radical}$ are applied to drive the reporter spins, one in each half of the NV spin echo sequence.  The change in magnetic field at the NV center caused by the reversal of the radical spins disrupts the NV spin echo, resulting in a dip in the NV coherence. (B) DEER spectrum as measured by a single NV center positioned below a photoactivated patch of dye/radical. Recorded at a constant applied field of 233 G, with $T_s$ fixed to 100 ns and $\tau =$ 900 ns. Blue curve is Lorentzian fit.}
\label{fig:Fig2}
\end{figure}

To probe the radical spins through their magnetic dipole interaction with the NV center, we implemented the DEER pulse sequence illustrated in Fig. \ref{fig:Fig2}A. The NV was first initialized into the $m_s = 0$ sublevel of its electronic ground state using a 532-nm optical pulse. A Hahn-echo pulse sequence was applied to the NV spin with a coherent microwave (MW) drive of frequency $\omega_{NV}$, resonant with the transition from $m_s=0$ to $m_s=-1$. Synchronously, we applied an RF drive of frequency $\omega_{radical} \neq \omega_{NV}$ to coherently manipulate the radical spins. Over the course of the measurement, the NV acquires phase in proportion to its interaction with the target spins. At the end of the sequence, the state of the NV is read out with a final optical pulse. All microscopy experiments were performed at an applied DC magnetic field around 200 G, aligned to the axis of the NV sensor. For the DEER measurements discussed below, we fixed the duration of the full sequence by setting $\tau = 900$ ns. Note that this is roughly fourfold shorter than the typical waiting time employed in DEER measurements, which is often set to coincide with the first revival of the spin-echo signal due to the NV's interaction with the surrounding $^{13}$C bath. Our choice to keep $\tau$ short helps to discriminate the signal due to converted dyes from the background of the more sparsely concentrated dark spins. The use of a radical drive pulse in both halves of the spin echo, and their temporal offset from the NV pulses, avoid certain spurious effects at the cost of introducing asymmetry \cite{mamin_detecting_2012,zhang_nanoscale_2021}. Either the frequency or the duration, $T_s$, of the radical drive can be swept. We quickly alternate measurements culminating in a final $\pi/2$ pulse on the NV with phases of $0^\circ$ or $180^\circ$ and take their difference to compute the resulting contrast. Normalizing by the fluorescence contrast measured in the absence of interaction results in a measurement of the NV coherence just before the last MW pulse (or more precisely, $-2\times$ the imaginary part of the coherence-- see Supporting Information).

The spectrum shown in Fig. \ref{fig:Fig2}B was recorded by fixing $T_s$ and sweeping $\omega_{radical}$. The center of the dip in NV coherence coincides with the expected resonance around 652 MHz at the applied field of 233 G. In this case, the linewidth of approximately 20 MHz is on the order of the hyperfine splittings of the radical anion formed from Alexa 488 as seen in high-resolution bulk EPR spectra \cite{van_de_linde_photoinduced_2011}.

\begin{figure*}[t]
\centering
\includegraphics[width=17.8cm,height=11.4cm]{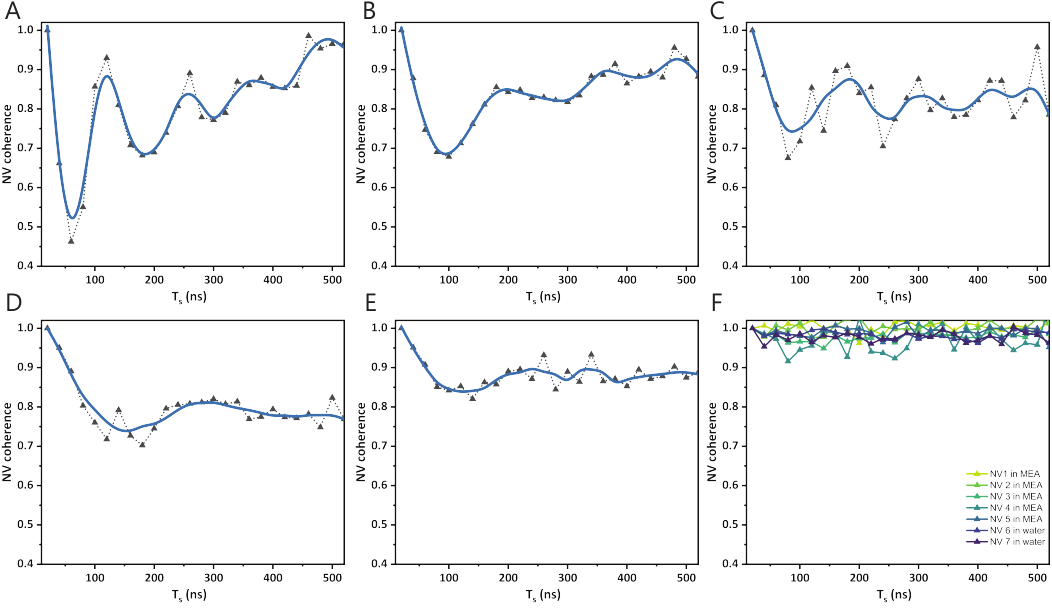}
\caption{Results of single-NV DEER measurements with frequency fixed to the Larmor frequency of the electron and $T_s$ varied. (A-D) Examples of DEER curves measured on covalently-functionalized diamond. (E) Example of DEER curve measured at diamond surface beneath a thin layer of PMMA containing Alexa Fluor 488. Additional data for this sample preparation are given in Fig. S9. In each of (A-E), the solid blue lines are intended to guide the eye and depict smoothed data averaged over a 5-point running window. (F) Results of control experiments on clean diamond surface.}
\label{fig:Fig3}
\end{figure*}

In the majority of our DEER measurements, we found it more informative to fix the RF frequency to the expected resonance and sweep $T_s$. Recordings from a number of different NVs are depicted in Fig. \ref{fig:Fig3}. In some cases we observed clear Rabi oscillations of the radical spins (Fig. \ref{fig:Fig3}A, B) that decay in amplitude as $T_s$ increases. Since $T_s \leq \tau = 900$ ns in our measurements, it is not valid to treat the radical spin drive as instantaneous, as is often done to simplify the analysis of DEER data \cite{mamin_detecting_2012,zhang_nanoscale_2021,dwyer_probing_2022}. In order to interpret the observed DEER curves, we found it necessary to include the finite $T_s$ directly in our theoretical modeling (Fig. S1 and accompanying text). In the Supporting Information we derive analytical results for two limiting cases: one in which the NV is coupled to many non-interacting quantum spins, and another in which the NV senses a classical magnetization whose dynamics are governed by Bloch equations with finite $T_1$ and $T_2$. Due to the finite $T_s$, the upward slant of the curves depicted in Fig. \ref{fig:Fig3}A, B is expected even in the absence of relaxation (Figs. S2-S4).

Figure \ref{fig:Fig3}C-E shows examples of another basic curve shape we repeatedly encountered, in which the oscillation appears highly damped, nearly plateauing at long $T_s$ to a value less than 1. Qualitatively similar shapes can be seen in cases where the target spins are strongly coupled to the NV center \cite{zhang_nanoscale_2021}, but our alternating measurement protocol allows us to rule this out since the plateaus occur well above zero NV coherence. The observed behavior does not appear to be consistent with an ensemble of non-interacting quantum spins (Fig. S3), nor of classical magnetization in the absence of relaxation (Fig. S4) or short $T_1$ (Fig. S5). Classical magnetization with a short $T_2$ does appear to yield similar overdamped curves (Fig. S6). To marry this observation with the quantum picture, we performed simulations of an NV coupled to a small ensemble of electronic spins, which in turn are coupled to one another via dipole-dipole interactions (see Supporting Information). Certain random realizations of the positions of the simulated target spins indeed result in qualitatively similar overdamped signals (Fig. S7). Moreover, different random realizations of the positions drawn from the same distribution can produce the full range of qualitative behaviors we observe in experiments, including cases showing more sustained oscillations with an upward-trending baseline, like those seen in experiment in Fig. \ref{fig:Fig3}A, B (Fig. S7).
As a set of controls, we removed the dye layer by cleaning the diamond in tri-acid mixture then piranha solution. We performed DEER measurements on shallow NVs within non-functionalized diamond, both in pure water and in the same MEA + pH 8 buffer used for our functionalized diamond experiments. Results from seven such measurements are displayed in Fig. \ref{fig:Fig3}F. With our truncated choice of $\tau$, any residual signal due to dark spins is clearly smaller in magnitude than those observed in the dye-coated diamond.

The complicated model needed to match the data makes it difficult to accurately extract a local radical spin density from the experimentally observed DEER curves. Nevertheless, defining even a simplistic estimator of this density is helpful for the sake of comparison. In the SI we detail conditions for which the observed DEER signal $\mathscr{S}$ obeys:
\begin{equation}
    \min_{T_s} \mathscr{S} \approx \exp\left[-\left(\frac{\mu_0}{4 \pi}\right)^2 \frac{3\pi\gamma^4 \hbar^2 \sigma}{16 d_{NV}^4} \tau^2 \right],
\end{equation}
where $\mu_0$ is the permeability of free space, $\gamma$ is the gyromagnetic ratio of an electron, $d_{NV}$ is the depth of the NV sensor below the diamond surface, and $\sigma$ is the areal density of target spins. Inverting this equation we arrive at our flawed-but-useful estimator of choice for the areal density:
\begin{equation}
    \hat{\sigma} \equiv - \ln{\left(\min_{T_s} \mathscr{S}\right)} \left(\frac{4\pi}{\mu_0}\right)^2 \left(\frac{16 \bar{d}^4_{NV}}{3\pi \gamma^4 \hbar^2 \tau^2}\right),
\end{equation}
where in lieu of measurements of the individual depth of each NV used in our experiments, we've replaced $d_{NV}$ by a constant mean NV depth $\bar{d}_{NV}$. We note that $\hat{\sigma}$ will tend to underestimate $\sigma$ when the effective $T_2$ of the target spins is short. Likewise it will tend to overestimate $\sigma$ if the true signal is below the noise floor. For the task of discriminating dye-derived spins from dark spins, therefore, $\hat{\sigma}$ is likely conservative. For each of the measurements in Fig. \ref{fig:Fig3}A-E, $\hat{\sigma}$ evaluates to 0.31, 0.16, 0.14, 0.14, and 0.10 spins per nm$^2$, respectively. For the seven controls depicted in Fig. \ref{fig:Fig3}F, $\hat{\sigma}$ gives 0.016, 0.0098, 0.016, 0.022, 0.020, 0.036, and 0.020. These numbers are generally consistent with previous estimates of dark spin concentrations on the surface of oxygen-annealed diamond \cite{dwyer_probing_2022, bluvstein_extending_2019,li_determination_2021}. On average, we find $\hat{\sigma}$ is about an order of magnitude larger for our measurements on dye-coated diamond as compared to the controls. Also notable is the fact that all $\hat{\sigma}$ values measured on dye-coated diamond are above the threshold of 0.05 per nm$^2$ presented as an upper bound for dark-spin density on similarly-prepared oxygen-terminated diamond surfaces in Ref. \cite{dwyer_probing_2022}. By contrast, $\hat{\sigma}$ for each control is below this threshold. While we cannot rule out the possibility that any one of our DEER signals observed on dye-coated diamond is due to an anomalously shallow NV that is strongly coupled to one or more anomalously stable dark spins, multiple pieces of evidence point to the conclusion that, at least on balance, we are indeed seeing magnetic signals due to the photoreduced dyes.

\subsection{Time-dependent measurements}

\begin{figure}[t]
\centering
\includegraphics[width=8.5cm,height=14.3cm]{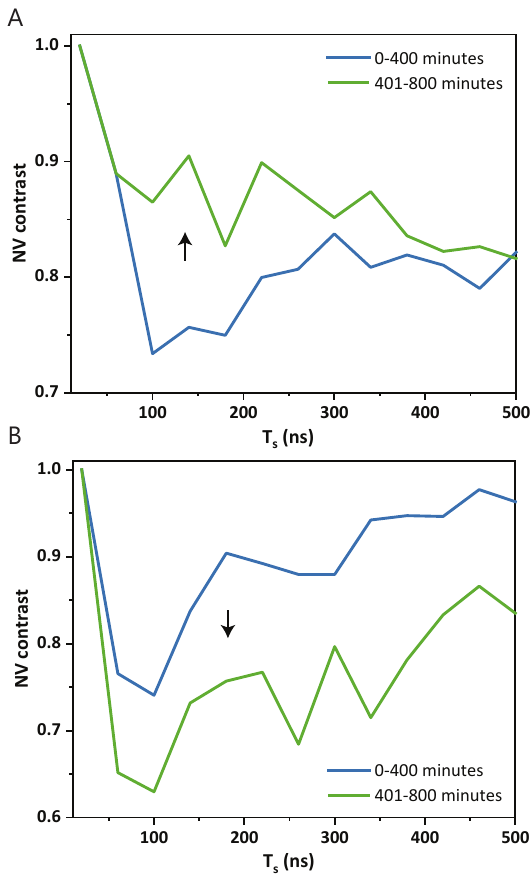}
\caption{Time-dependent DEER measurements. (A) A loss of DEER signal amplitude, likely due to conversion of the radicals back to the fluorescent state. Data are from the same measurement as in Fig. \ref{fig:Fig3}D. (B) An apparent gain of DEER signal amplitude over time, perhaps due to prolonged exposure to 532-nm laser used to initialize and readout NV. Data are from same measurement as in Fig. \ref{fig:Fig3}B.}
\label{fig:Fig4}
\end{figure}

Bulk EPR data show that as the radicals convert back to the fluorescent state, the magnetic signal decreases over time after the initiating laser pulse \cite{van_de_linde_photoinduced_2011} (Fig. \ref{fig:Fig1}B). We generally observed that holes burned into the surface-attached dye ensemble recovered their fluorescence after being left in the dark for a few hours. In Fig. \ref{fig:Fig4}A we replot some of the data from the same single-NV measurement as in Fig. \ref{fig:Fig3}D. The overnight-averaged data are split into those acquired in the first 400 minutes vs. the second 400 minutes after photoactivation. Data from neighboring pairs of $T_s$ values are averaged together to compensate the accompanying decrease in signal-to-noise ratio. Between the two measurement periods, The extracted value of $\hat{\sigma}$ is depressed from 0.13 to 0.08 per nm$^2$. This is in qualitative accord with the trend observed in bulk measurements, though disagreement in the exact timescale of conversion from magnetic to fluorescent state are to be expected due to differences in the concentrations of reducing agent and dye, as well as the fundamentally different nanoscale environment at the diamond surface. In other cases, we observed time-dependent behavior that was less straightforward to rationalize. In Fig. \ref{fig:Fig4}B we replot data from the same measurement as in Fig. \ref{fig:Fig3}B, again splitting into two sequential periods of time. Here the NV coherence normalized to the contrast at the first time point ($T_s=$20 ns) decreases over time, commensurate with an increase in $\hat{\sigma}$ from 0.12 to 0.19 per nm$^2$. We hypothesize that the local concentration of radicals might increase over time in this case due to the continual exposure to 532-nm light intended to address the NV. The absorption spectrum of Alexa Fluor 488 is about 5\% of its maximum at 532 nm (source: AAT Bioquest website), which isn't entirely insignificant given the many hours spent repeating the measurement sequence. The direction in which the pseudo-equilibrium between fluorescent dye and radical anion is pushed over the course of the measurement might depend intricately on local concentrations and laser intensities. Interpretation is further complicated by the fact that for the data depicted in Fig. \ref{fig:Fig4}B, the absolute contrast (as opposed to that normalized to $T_s=$ 20 ns) decreases by more than half from Period 1 to Period 2.


\section{CONCLUSIONS}

In summary, we have demonstrated that by interfacing with single, shallow NV centers for read out, radicals formed from photoreduced fluorescent dyes can be wielded as reporters of their nanoscale magnetic environment. At the densities realized in this work, the nanoscale magnetic environment of any one such radical appears to be largely determined by couplings to other radicals in its vicinity. Different random configurations of such a qubit network can account for the observed heterogeneity in line shapes from site to site. In the future, we aim to interface single NVs with single dye-derived radicals, which might provide a more general route to sensing single external nuclear spins than relying on stable dark spins \cite{sushkov_all-optical_2014}. In the current work, we make use of a roughly tenfold disparity in areal density to help distinguish the observed signal from dark-spin background. At lower dye densities, other strategies like optical co-localization will be needed to help make this distinction. In future work, the density of dark spins can be suppressed with a thin dielectric coating deposited by atomic layer deposition (ALD) \cite{yu_engineering_2025}. Such an ALD layer might also help insulate the NVs from the deleterious effects of the elevated pH needed to activate MEA as a reducing agent, though the use of alternative mild reducing agents could extend the operating pH range as well. In addition to being wielded as reporters from the exterior of the diamond to the NVs, surface-localized radicals derived from dye molecules could also be used to transfer polarization from NVs to external nuclei for improved sensing \cite{espinos_enhancing_2024, belthangady_dressed-state_2013}.

The prospect of correlative super-resolution fluorescence and nanoscale magnetic resonance imaging constitutes another exciting avenue for future work. Powerful biological insights have been gleaned from experiments in which super-resolution fluorescence microscopy is annotated by a complementary imaging technique, such as cryo-electron tomography \cite{dahlberg2021cryogenic}. Here we have the opportunity to annotate with a different complementary imaging modality, where interchangeable forms of the same labeling molecules can be used to alternately generate optical or magnetic contrast.

Dense ensembles of surface-attached photoswitchable radicals might also be an attractive platform for quantum many-body simulation on low-dimensional, disordered networks \cite{davis2023probing}, or for entanglement-enhanced sensing \cite{HughesPRX2025}. Custom-designed 1D and 2D networks of disordered spins can be lithographically printed by selective exposure to the photoswitching laser. In principle this could be done with subdiffraction resolution by adapting something like stimulated emission depletion (STED) microscopy \cite{hell1994breaking}. More regular networks of such spins could be synthesized using DNA origami \cite{schmied2012fluorescence}.

Finally, future work will explore the suitability of photoreduced fluorescent dyes as sensing qubits unto themselves, without the constraint of proximity to an NV. Freeing the dye-derived radicals from the diamond surface would enable a much wider range of in situ sensing possibilities, so long as there is some intrinsic spin-optical phenomenon to take advantage of for initialization and read out. A flurry of recent studies have presented new candidates for molecular qubits, especially for sensing \cite{bayliss_laorenza_mintun_kovos_freedman_awschalom_2020,kopp_nakamura_phelan_poh_tyndall_brown_huang_yuen-zhou_krzyaniak_wasielewski_2024, chowdhury_murto_panjwani_sun_ghosh_boeije_derkach_woo_millington_congrave_etal._2024, sutcliffe_kazmierczak_hadt_2024,zhou_sun_sun_2024,feder2025fluorescent, meng_nie_berger_nick_einholz_rizzato_schleicher_bucher_2025, mann_chemically_2025}. Determining whether radicals formed from photoreduced fluorescent dyes might constitute a competitive molecular qubit technology requires a deeper understanding of the radical's photophysics.

\begin{acknowledgments}
We thank Dr. Toby Woods for support with the cw-EPR experiments and Dr. Austin Cyphersmith for assistance with imaging dye functionalized diamond at Carl R. Woese Institute for Genomic Biology. This work was supported in part by the Arnold and Mabel Beckman Foundation (via a Beckman Young Investigator Award to M.P.B.) and by the Spectroscopy Society of Pittsburgh (via a Starter Grant Award to M.P.B). L.P.A. and A.N.R. acknowledge  support from the Thor R. Rubin Fellowship. L.P.A. acknowledges additional support from the Lester E. and Kathleen A. Coleman Fellowship, and the Victor E. Buhrke Graduate Fellowship. We acknowledge the use of Materials Research Laboratory annealing furnace facility at University of Illinois Urbana-Champaign
\end{acknowledgments}

\appendix

\section{Diamond Sample}
An electronic-grade diamond substrate (Element Six) was implanted with $^{15}$N ions (Coherent-Innovion) at 7 keV and a 7$^\circ$ tilt. One half of the $4 \mathrm{mm} \times 4 \mathrm{mm} \times 0.5 \mathrm{mm}$ chip was implanted at a fluence of $1\times10^{9} \mathrm{cm}^{-2}$ to generate isolated substitutional nitrogen atoms for single NV measurements. The other half was implanted at a higher fluence of $3\times10^{13} \mathrm{cm}^{-2}$ to produce ensembles. To prepare the NVs, the diamond was subsequently annealed in vacuum at 800$^\circ$C  in our custom built vacuum annealer. The diamond was then washed in piranha solution, oxygen annealed and then washed again in piranha. Just before dye functionalizing, the diamond was cleaned in an equal parts mixture of nitric, perchloric, and sulfuric acid.

\section{Dye Functionalization}
Alexa Fluor 488 dye molecules were covalently attached to an oxygen-annealed diamond chip using carbodiimide crosslinker chemistry. Carboxylic acid groups on the diamond surface were first conjugated to ethylenediamine via 1-ethyl-3-(3-dimethylaminopropyl)carbodiimide hydrochloride (EDC) chemistry, with N-hydroxysuccinimide (NHS) included in the reaction mixture to enhance coupling efficiency. The resulting amine terminated surface was then reacted with Alexa Fluor 488 succinimidyl ester to complete the functionalization. The reagents EDC, NHS, ethylenediamine, and Alexa Fluor 488 NHS ester were purchased from ThermoFisher Scientific (77149, 24500, A12132.0F, and A20000, respectively). For photoinduced reduction, the weak reducing agent $\beta$-mercaptoethylamine (MEA) (Fisher Scientific, AAA1437714) was dissolved in deionized water, and the solution pH was adjusted using 10 N NaOH (Sigma Aldrich, 1310-73-2). The functionalized diamond chip was mounted onto a bottomless channel slide (ibidi sticky-Slide I Luer 0.6 mm, 80168) and sealed with a 24 mm $\times$ 54 mm coverslip (ibidi, 10832). MEA solution was then pipetted into the channel slide. Additional details on the reaction, as well as on an alternative dye-coating strategy employed for some experiments can be found in the Supporting Information.

\section{Experimental Setup}
Experiments were performed on a custom-built confocal microscope based on a PInano Stage Platform System (Physik Instrumente). A 532-nm laser with an intensity of approximately 1 MW/cm$^2$ (Coherent, Sapphire 532-200 CW CDRH USB Laser System) was used to polarize and read out NV centers within a diffraction-limited spot. The laser was gated with an acousto-optic modulator (GH AOM, model 3250-220) and focused onto the diamond using a 100$\times$ NA 1.45 oil-immersion objective (Olympus, UPLXAPO100X). NV fluorescence was collected through the same objective and passed through a dichroic mirror (Semrock, Di03-R532-t1-25\mbox{$\times$}36), a long-pass filter (Semrock, BLP01-635R-25), and a 532-nm notch filter (Semrock, NF01-532U-25) before detection on a fiber-coupled single-photon-counting module (Excelitas, SPCM-AQRH-14-FC-ND). Data acquisition was performed using an NI USB-6361, X Series DAQ. NVs were coherently driven at Rabi frequencies in the range of 10-25 MHz by microwaves generated from an SRS SG384 with built-in IQ modulation. The microwaves were gated using a high-speed switch (Mini-Circuits, ZASWA2-50DR-FA), then amplified (Mini-Circuits, ZHL-16W-43-S+). Radical spins were driven using RF generated either by a Windfreak SynthNV signal generator or a second SRS SG384, amplified (ZHL-15W-422-S+), and combined with the NV microwave signal using a two way power combiner (ZAPD-30-S+), then delivered to the diamond surface through a 44-AWG magnet wire (Remington Industries). A neodymium magnet (KJ Magnetics) mounted near the diamond was used to apply a bias magnetic field of $\sim$200 G.

A 488-nm laser (Coherent, Sapphire 488-200 CW CDRH USB Laser System), added to the excitation path was used to generate dye radicals. Dye solution measurements were done after continuous photoexcitation for 5 minutes with peak intensity $~$25 MW/cm$^2$ distributed over a 3D volume of solution. Surface-bound dye molecules were excited at approximately $5 \times{10}^4$ W/cm$^2$ for 30 - 120 seconds to generate radicals for the DEER measurements.

\section{Theoretical modeling and simulation}
Details on analytical modeling and numerical simulation can be found in the Supporting Information.


\bibliography{References}

\begin{thebibliography}{61}%
\makeatletter
\providecommand \@ifxundefined [1]{%
 \@ifx{#1\undefined}
}%
\providecommand \@ifnum [1]{%
 \ifnum #1\expandafter \@firstoftwo
 \else \expandafter \@secondoftwo
 \fi
}%
\providecommand \@ifx [1]{%
 \ifx #1\expandafter \@firstoftwo
 \else \expandafter \@secondoftwo
 \fi
}%
\providecommand \natexlab [1]{#1}%
\providecommand \enquote  [1]{``#1''}%
\providecommand \bibnamefont  [1]{#1}%
\providecommand \bibfnamefont [1]{#1}%
\providecommand \citenamefont [1]{#1}%
\providecommand \href@noop [0]{\@secondoftwo}%
\providecommand \href [0]{\begingroup \@sanitize@url \@href}%
\providecommand \@href[1]{\@@startlink{#1}\@@href}%
\providecommand \@@href[1]{\endgroup#1\@@endlink}%
\providecommand \@sanitize@url [0]{\catcode `\\12\catcode `\$12\catcode `\&12\catcode `\#12\catcode `\^12\catcode `\_12\catcode `\%12\relax}%
\providecommand \@@startlink[1]{}%
\providecommand \@@endlink[0]{}%
\providecommand \url  [0]{\begingroup\@sanitize@url \@url }%
\providecommand \@url [1]{\endgroup\@href {#1}{\urlprefix }}%
\providecommand \urlprefix  [0]{URL }%
\providecommand \Eprint [0]{\href }%
\providecommand \doibase [0]{https://doi.org/}%
\providecommand \selectlanguage [0]{\@gobble}%
\providecommand \bibinfo  [0]{\@secondoftwo}%
\providecommand \bibfield  [0]{\@secondoftwo}%
\providecommand \translation [1]{[#1]}%
\providecommand \BibitemOpen [0]{}%
\providecommand \bibitemStop [0]{}%
\providecommand \bibitemNoStop [0]{.\EOS\space}%
\providecommand \EOS [0]{\spacefactor3000\relax}%
\providecommand \BibitemShut  [1]{\csname bibitem#1\endcsname}%
\let\auto@bib@innerbib\@empty
\bibitem [{\citenamefont {Jelezko}\ \emph {et~al.}(2004)\citenamefont {Jelezko}, \citenamefont {Gaebel}, \citenamefont {Popa}, \citenamefont {Gruber},\ and\ \citenamefont {Wrachtrup}}]{jelezko_observation_2004}%
  \BibitemOpen
  \bibfield  {author} {\bibinfo {author} {\bibfnamefont {F.}~\bibnamefont {Jelezko}}, \bibinfo {author} {\bibfnamefont {T.}~\bibnamefont {Gaebel}}, \bibinfo {author} {\bibfnamefont {I.}~\bibnamefont {Popa}}, \bibinfo {author} {\bibfnamefont {A.}~\bibnamefont {Gruber}},\ and\ \bibinfo {author} {\bibfnamefont {J.}~\bibnamefont {Wrachtrup}},\ }\bibfield  {title} {\bibinfo {title} {Observation of {Coherent} {Oscillations} in a {Single} {Electron} {Spin}},\ }\href {https://doi.org/10.1103/PhysRevLett.92.076401} {\bibfield  {journal} {\bibinfo  {journal} {Physical Review Letters}\ }\textbf {\bibinfo {volume} {92}},\ \bibinfo {pages} {076401} (\bibinfo {year} {2004})}\BibitemShut {NoStop}%
\bibitem [{\citenamefont {Schirhagl}\ \emph {et~al.}(2014)\citenamefont {Schirhagl}, \citenamefont {Chang}, \citenamefont {Loretz},\ and\ \citenamefont {Degen}}]{schirhagl_nitrogen-vacancy_2014}%
  \BibitemOpen
  \bibfield  {author} {\bibinfo {author} {\bibfnamefont {R.}~\bibnamefont {Schirhagl}}, \bibinfo {author} {\bibfnamefont {K.}~\bibnamefont {Chang}}, \bibinfo {author} {\bibfnamefont {M.}~\bibnamefont {Loretz}},\ and\ \bibinfo {author} {\bibfnamefont {C.~L.}\ \bibnamefont {Degen}},\ }\bibfield  {title} {\bibinfo {title} {Nitrogen-{Vacancy} {Centers} in {Diamond}: {Nanoscale} {Sensors} for {Physics} and {Biology}},\ }\href {https://doi.org/https://doi.org/10.1146/annurev-physchem-040513-103659} {\bibfield  {journal} {\bibinfo  {journal} {Annual Review of Physical Chemistry}\ }\textbf {\bibinfo {volume} {65}},\ \bibinfo {pages} {83} (\bibinfo {year} {2014})},\ \bibinfo {note} {publisher: Annual Reviews Type: Journal Article}\BibitemShut {NoStop}%
\bibitem [{\citenamefont {Heremans}\ \emph {et~al.}(2016)\citenamefont {Heremans}, \citenamefont {Yale},\ and\ \citenamefont {Awschalom}}]{heremans_control_2016}%
  \BibitemOpen
  \bibfield  {author} {\bibinfo {author} {\bibfnamefont {F.~J.}\ \bibnamefont {Heremans}}, \bibinfo {author} {\bibfnamefont {C.~G.}\ \bibnamefont {Yale}},\ and\ \bibinfo {author} {\bibfnamefont {D.~D.}\ \bibnamefont {Awschalom}},\ }\bibfield  {title} {\bibinfo {title} {Control of {Spin} {Defects} in {Wide}-{Bandgap} {Semiconductors} for {Quantum} {Technologies}},\ }\href {https://doi.org/10.1109/JPROC.2016.2561274} {\bibfield  {journal} {\bibinfo  {journal} {Proceedings of the IEEE}\ }\textbf {\bibinfo {volume} {104}},\ \bibinfo {pages} {2009} (\bibinfo {year} {2016})}\BibitemShut {NoStop}%
\bibitem [{\citenamefont {Taylor}\ \emph {et~al.}(2008)\citenamefont {Taylor}, \citenamefont {Cappellaro}, \citenamefont {Childress}, \citenamefont {Jiang}, \citenamefont {Budker}, \citenamefont {Hemmer}, \citenamefont {Yacoby}, \citenamefont {Walsworth},\ and\ \citenamefont {Lukin}}]{taylor_high-sensitivity_2008}%
  \BibitemOpen
  \bibfield  {author} {\bibinfo {author} {\bibfnamefont {J.~M.}\ \bibnamefont {Taylor}}, \bibinfo {author} {\bibfnamefont {P.}~\bibnamefont {Cappellaro}}, \bibinfo {author} {\bibfnamefont {L.}~\bibnamefont {Childress}}, \bibinfo {author} {\bibfnamefont {L.}~\bibnamefont {Jiang}}, \bibinfo {author} {\bibfnamefont {D.}~\bibnamefont {Budker}}, \bibinfo {author} {\bibfnamefont {P.~R.}\ \bibnamefont {Hemmer}}, \bibinfo {author} {\bibfnamefont {A.}~\bibnamefont {Yacoby}}, \bibinfo {author} {\bibfnamefont {R.}~\bibnamefont {Walsworth}},\ and\ \bibinfo {author} {\bibfnamefont {M.~D.}\ \bibnamefont {Lukin}},\ }\bibfield  {title} {\bibinfo {title} {High-sensitivity diamond magnetometer with nanoscale resolution},\ }\href {https://doi.org/10.1038/nphys1075} {\bibfield  {journal} {\bibinfo  {journal} {Nature Physics}\ }\textbf {\bibinfo {volume} {4}},\ \bibinfo {pages} {810} (\bibinfo {year} {2008})}\BibitemShut {NoStop}%
\bibitem [{\citenamefont {Müller}\ \emph {et~al.}(2014)\citenamefont {Müller}, \citenamefont {Kong}, \citenamefont {Cai}, \citenamefont {Melentijević}, \citenamefont {Stacey}, \citenamefont {Markham}, \citenamefont {Twitchen}, \citenamefont {Isoya}, \citenamefont {Pezzagna}, \citenamefont {Meijer}, \citenamefont {Du}, \citenamefont {Plenio}, \citenamefont {Naydenov}, \citenamefont {McGuinness},\ and\ \citenamefont {Jelezko}}]{muller_nuclear_2014}%
  \BibitemOpen
  \bibfield  {author} {\bibinfo {author} {\bibfnamefont {C.}~\bibnamefont {Müller}}, \bibinfo {author} {\bibfnamefont {X.}~\bibnamefont {Kong}}, \bibinfo {author} {\bibfnamefont {J.-M.}\ \bibnamefont {Cai}}, \bibinfo {author} {\bibfnamefont {K.}~\bibnamefont {Melentijević}}, \bibinfo {author} {\bibfnamefont {A.}~\bibnamefont {Stacey}}, \bibinfo {author} {\bibfnamefont {M.}~\bibnamefont {Markham}}, \bibinfo {author} {\bibfnamefont {D.}~\bibnamefont {Twitchen}}, \bibinfo {author} {\bibfnamefont {J.}~\bibnamefont {Isoya}}, \bibinfo {author} {\bibfnamefont {S.}~\bibnamefont {Pezzagna}}, \bibinfo {author} {\bibfnamefont {J.}~\bibnamefont {Meijer}}, \bibinfo {author} {\bibfnamefont {J.~F.}\ \bibnamefont {Du}}, \bibinfo {author} {\bibfnamefont {M.~B.}\ \bibnamefont {Plenio}}, \bibinfo {author} {\bibfnamefont {B.}~\bibnamefont {Naydenov}}, \bibinfo {author} {\bibfnamefont {L.~P.}\ \bibnamefont {McGuinness}},\ and\ \bibinfo {author} {\bibfnamefont {F.}~\bibnamefont {Jelezko}},\ }\bibfield  {title} {\bibinfo {title}
  {Nuclear magnetic resonance spectroscopy with single spin sensitivity},\ }\href {https://doi.org/10.1038/ncomms5703} {\bibfield  {journal} {\bibinfo  {journal} {Nature Communications}\ }\textbf {\bibinfo {volume} {5}},\ \bibinfo {pages} {4703} (\bibinfo {year} {2014})}\BibitemShut {NoStop}%
\bibitem [{\citenamefont {Zopes}\ \emph {et~al.}(2018)\citenamefont {Zopes}, \citenamefont {Cujia}, \citenamefont {Sasaki}, \citenamefont {Boss}, \citenamefont {Itoh},\ and\ \citenamefont {Degen}}]{zopes_three-dimensional_2018}%
  \BibitemOpen
  \bibfield  {author} {\bibinfo {author} {\bibfnamefont {J.}~\bibnamefont {Zopes}}, \bibinfo {author} {\bibfnamefont {K.~S.}\ \bibnamefont {Cujia}}, \bibinfo {author} {\bibfnamefont {K.}~\bibnamefont {Sasaki}}, \bibinfo {author} {\bibfnamefont {J.~M.}\ \bibnamefont {Boss}}, \bibinfo {author} {\bibfnamefont {K.~M.}\ \bibnamefont {Itoh}},\ and\ \bibinfo {author} {\bibfnamefont {C.~L.}\ \bibnamefont {Degen}},\ }\bibfield  {title} {\bibinfo {title} {Three-dimensional localization spectroscopy of individual nuclear spins with sub-{Angstrom} resolution},\ }\href {https://doi.org/10.1038/s41467-018-07121-0} {\bibfield  {journal} {\bibinfo  {journal} {Nature Communications}\ }\textbf {\bibinfo {volume} {9}},\ \bibinfo {pages} {4678} (\bibinfo {year} {2018})}\BibitemShut {NoStop}%
\bibitem [{\citenamefont {Maze}\ \emph {et~al.}(2008)\citenamefont {Maze}, \citenamefont {Stanwix}, \citenamefont {Hodges}, \citenamefont {Hong}, \citenamefont {Taylor}, \citenamefont {Cappellaro}, \citenamefont {Jiang}, \citenamefont {Dutt}, \citenamefont {Togan}, \citenamefont {Zibrov}, \citenamefont {Yacoby}, \citenamefont {Walsworth},\ and\ \citenamefont {Lukin}}]{maze_nanoscale_2008}%
  \BibitemOpen
  \bibfield  {author} {\bibinfo {author} {\bibfnamefont {J.~R.}\ \bibnamefont {Maze}}, \bibinfo {author} {\bibfnamefont {P.~L.}\ \bibnamefont {Stanwix}}, \bibinfo {author} {\bibfnamefont {J.~S.}\ \bibnamefont {Hodges}}, \bibinfo {author} {\bibfnamefont {S.}~\bibnamefont {Hong}}, \bibinfo {author} {\bibfnamefont {J.~M.}\ \bibnamefont {Taylor}}, \bibinfo {author} {\bibfnamefont {P.}~\bibnamefont {Cappellaro}}, \bibinfo {author} {\bibfnamefont {L.}~\bibnamefont {Jiang}}, \bibinfo {author} {\bibfnamefont {M.~V.~G.}\ \bibnamefont {Dutt}}, \bibinfo {author} {\bibfnamefont {E.}~\bibnamefont {Togan}}, \bibinfo {author} {\bibfnamefont {A.~S.}\ \bibnamefont {Zibrov}}, \bibinfo {author} {\bibfnamefont {A.}~\bibnamefont {Yacoby}}, \bibinfo {author} {\bibfnamefont {R.~L.}\ \bibnamefont {Walsworth}},\ and\ \bibinfo {author} {\bibfnamefont {M.~D.}\ \bibnamefont {Lukin}},\ }\bibfield  {title} {\bibinfo {title} {Nanoscale magnetic sensing with an individual electronic spin in diamond},\ }\href
  {https://doi.org/10.1038/nature07279} {\bibfield  {journal} {\bibinfo  {journal} {Nature}\ }\textbf {\bibinfo {volume} {455}},\ \bibinfo {pages} {644} (\bibinfo {year} {2008})}\BibitemShut {NoStop}%
\bibitem [{\citenamefont {Grinolds}\ \emph {et~al.}(2014)\citenamefont {Grinolds}, \citenamefont {Warner}, \citenamefont {De~Greve}, \citenamefont {Dovzhenko}, \citenamefont {Thiel}, \citenamefont {Walsworth}, \citenamefont {Hong}, \citenamefont {Maletinsky},\ and\ \citenamefont {Yacoby}}]{grinolds_subnanometre_2014}%
  \BibitemOpen
  \bibfield  {author} {\bibinfo {author} {\bibfnamefont {M.~S.}\ \bibnamefont {Grinolds}}, \bibinfo {author} {\bibfnamefont {M.}~\bibnamefont {Warner}}, \bibinfo {author} {\bibfnamefont {K.}~\bibnamefont {De~Greve}}, \bibinfo {author} {\bibfnamefont {Y.}~\bibnamefont {Dovzhenko}}, \bibinfo {author} {\bibfnamefont {L.}~\bibnamefont {Thiel}}, \bibinfo {author} {\bibfnamefont {R.~L.}\ \bibnamefont {Walsworth}}, \bibinfo {author} {\bibfnamefont {S.}~\bibnamefont {Hong}}, \bibinfo {author} {\bibfnamefont {P.}~\bibnamefont {Maletinsky}},\ and\ \bibinfo {author} {\bibfnamefont {A.}~\bibnamefont {Yacoby}},\ }\bibfield  {title} {\bibinfo {title} {Subnanometre resolution in three-dimensional magnetic resonance imaging of individual dark spins},\ }\href {https://doi.org/10.1038/nnano.2014.30} {\bibfield  {journal} {\bibinfo  {journal} {Nature Nanotechnology}\ }\textbf {\bibinfo {volume} {9}},\ \bibinfo {pages} {279} (\bibinfo {year} {2014})}\BibitemShut {NoStop}%
\bibitem [{\citenamefont {DeVience}\ \emph {et~al.}(2015)\citenamefont {DeVience}, \citenamefont {Pham}, \citenamefont {Lovchinsky}, \citenamefont {Sushkov}, \citenamefont {Bar-Gill}, \citenamefont {Belthangady}, \citenamefont {Casola}, \citenamefont {Corbett}, \citenamefont {Zhang}, \citenamefont {Lukin}, \citenamefont {Park}, \citenamefont {Yacoby},\ and\ \citenamefont {Walsworth}}]{devience_nanoscale_2015}%
  \BibitemOpen
  \bibfield  {author} {\bibinfo {author} {\bibfnamefont {S.~J.}\ \bibnamefont {DeVience}}, \bibinfo {author} {\bibfnamefont {L.~M.}\ \bibnamefont {Pham}}, \bibinfo {author} {\bibfnamefont {I.}~\bibnamefont {Lovchinsky}}, \bibinfo {author} {\bibfnamefont {A.~O.}\ \bibnamefont {Sushkov}}, \bibinfo {author} {\bibfnamefont {N.}~\bibnamefont {Bar-Gill}}, \bibinfo {author} {\bibfnamefont {C.}~\bibnamefont {Belthangady}}, \bibinfo {author} {\bibfnamefont {F.}~\bibnamefont {Casola}}, \bibinfo {author} {\bibfnamefont {M.}~\bibnamefont {Corbett}}, \bibinfo {author} {\bibfnamefont {H.}~\bibnamefont {Zhang}}, \bibinfo {author} {\bibfnamefont {M.}~\bibnamefont {Lukin}}, \bibinfo {author} {\bibfnamefont {H.}~\bibnamefont {Park}}, \bibinfo {author} {\bibfnamefont {A.}~\bibnamefont {Yacoby}},\ and\ \bibinfo {author} {\bibfnamefont {R.~L.}\ \bibnamefont {Walsworth}},\ }\bibfield  {title} {\bibinfo {title} {Nanoscale {NMR} spectroscopy and imaging of multiple nuclear species},\ }\href {https://doi.org/10.1038/nnano.2014.313}
  {\bibfield  {journal} {\bibinfo  {journal} {Nature Nanotechnology}\ }\textbf {\bibinfo {volume} {10}},\ \bibinfo {pages} {129} (\bibinfo {year} {2015})}\BibitemShut {NoStop}%
\bibitem [{\citenamefont {Abobeih}\ \emph {et~al.}(2019)\citenamefont {Abobeih}, \citenamefont {Randall}, \citenamefont {Bradley}, \citenamefont {Bartling}, \citenamefont {Bakker}, \citenamefont {Degen}, \citenamefont {Markham}, \citenamefont {Twitchen},\ and\ \citenamefont {Taminiau}}]{abobeih_atomic-scale_2019}%
  \BibitemOpen
  \bibfield  {author} {\bibinfo {author} {\bibfnamefont {M.~H.}\ \bibnamefont {Abobeih}}, \bibinfo {author} {\bibfnamefont {J.}~\bibnamefont {Randall}}, \bibinfo {author} {\bibfnamefont {C.~E.}\ \bibnamefont {Bradley}}, \bibinfo {author} {\bibfnamefont {H.~P.}\ \bibnamefont {Bartling}}, \bibinfo {author} {\bibfnamefont {M.~A.}\ \bibnamefont {Bakker}}, \bibinfo {author} {\bibfnamefont {M.~J.}\ \bibnamefont {Degen}}, \bibinfo {author} {\bibfnamefont {M.}~\bibnamefont {Markham}}, \bibinfo {author} {\bibfnamefont {D.~J.}\ \bibnamefont {Twitchen}},\ and\ \bibinfo {author} {\bibfnamefont {T.~H.}\ \bibnamefont {Taminiau}},\ }\bibfield  {title} {\bibinfo {title} {Atomic-scale imaging of a 27-nuclear-spin cluster using a quantum sensor},\ }\href {https://doi.org/10.1038/s41586-019-1834-7} {\bibfield  {journal} {\bibinfo  {journal} {Nature}\ }\textbf {\bibinfo {volume} {576}},\ \bibinfo {pages} {411} (\bibinfo {year} {2019})}\BibitemShut {NoStop}%
\bibitem [{\citenamefont {Balasubramanian}\ \emph {et~al.}(2008)\citenamefont {Balasubramanian}, \citenamefont {Chan}, \citenamefont {Kolesov}, \citenamefont {Al-Hmoud}, \citenamefont {Tisler}, \citenamefont {Shin}, \citenamefont {Kim}, \citenamefont {Wojcik}, \citenamefont {Hemmer}, \citenamefont {Krueger}, \citenamefont {Hanke}, \citenamefont {Leitenstorfer}, \citenamefont {Bratschitsch}, \citenamefont {Jelezko},\ and\ \citenamefont {Wrachtrup}}]{balasubramanian_nanoscale_2008}%
  \BibitemOpen
  \bibfield  {author} {\bibinfo {author} {\bibfnamefont {G.}~\bibnamefont {Balasubramanian}}, \bibinfo {author} {\bibfnamefont {I.~Y.}\ \bibnamefont {Chan}}, \bibinfo {author} {\bibfnamefont {R.}~\bibnamefont {Kolesov}}, \bibinfo {author} {\bibfnamefont {M.}~\bibnamefont {Al-Hmoud}}, \bibinfo {author} {\bibfnamefont {J.}~\bibnamefont {Tisler}}, \bibinfo {author} {\bibfnamefont {C.}~\bibnamefont {Shin}}, \bibinfo {author} {\bibfnamefont {C.}~\bibnamefont {Kim}}, \bibinfo {author} {\bibfnamefont {A.}~\bibnamefont {Wojcik}}, \bibinfo {author} {\bibfnamefont {P.~R.}\ \bibnamefont {Hemmer}}, \bibinfo {author} {\bibfnamefont {A.}~\bibnamefont {Krueger}}, \bibinfo {author} {\bibfnamefont {T.}~\bibnamefont {Hanke}}, \bibinfo {author} {\bibfnamefont {A.}~\bibnamefont {Leitenstorfer}}, \bibinfo {author} {\bibfnamefont {R.}~\bibnamefont {Bratschitsch}}, \bibinfo {author} {\bibfnamefont {F.}~\bibnamefont {Jelezko}},\ and\ \bibinfo {author} {\bibfnamefont {J.}~\bibnamefont {Wrachtrup}},\ }\bibfield  {title} {\bibinfo
  {title} {Nanoscale imaging magnetometry with diamond spins under ambient conditions},\ }\href {https://doi.org/10.1038/nature07278} {\bibfield  {journal} {\bibinfo  {journal} {Nature}\ }\textbf {\bibinfo {volume} {455}},\ \bibinfo {pages} {648} (\bibinfo {year} {2008})}\BibitemShut {NoStop}%
\bibitem [{\citenamefont {Lovchinsky}\ \emph {et~al.}(2016)\citenamefont {Lovchinsky}, \citenamefont {Sushkov}, \citenamefont {Urbach}, \citenamefont {Leon}, \citenamefont {Choi}, \citenamefont {Greve}, \citenamefont {Evans}, \citenamefont {Gertner}, \citenamefont {Bersin}, \citenamefont {Müller}, \citenamefont {McGuinness}, \citenamefont {Jelezko}, \citenamefont {Walsworth}, \citenamefont {Park},\ and\ \citenamefont {Lukin}}]{lovchinsky_nuclear_2016}%
  \BibitemOpen
  \bibfield  {author} {\bibinfo {author} {\bibfnamefont {I.}~\bibnamefont {Lovchinsky}}, \bibinfo {author} {\bibfnamefont {A.~O.}\ \bibnamefont {Sushkov}}, \bibinfo {author} {\bibfnamefont {E.}~\bibnamefont {Urbach}}, \bibinfo {author} {\bibfnamefont {N.~P.~d.}\ \bibnamefont {Leon}}, \bibinfo {author} {\bibfnamefont {S.}~\bibnamefont {Choi}}, \bibinfo {author} {\bibfnamefont {K.~D.}\ \bibnamefont {Greve}}, \bibinfo {author} {\bibfnamefont {R.}~\bibnamefont {Evans}}, \bibinfo {author} {\bibfnamefont {R.}~\bibnamefont {Gertner}}, \bibinfo {author} {\bibfnamefont {E.}~\bibnamefont {Bersin}}, \bibinfo {author} {\bibfnamefont {C.}~\bibnamefont {Müller}}, \bibinfo {author} {\bibfnamefont {L.}~\bibnamefont {McGuinness}}, \bibinfo {author} {\bibfnamefont {F.}~\bibnamefont {Jelezko}}, \bibinfo {author} {\bibfnamefont {R.~L.}\ \bibnamefont {Walsworth}}, \bibinfo {author} {\bibfnamefont {H.}~\bibnamefont {Park}},\ and\ \bibinfo {author} {\bibfnamefont {M.~D.}\ \bibnamefont {Lukin}},\ }\bibfield  {title} {\bibinfo
  {title} {Nuclear magnetic resonance detection and spectroscopy of single proteins using quantum logic},\ }\href {https://doi.org/10.1126/science.aad8022} {\bibfield  {journal} {\bibinfo  {journal} {Science}\ }\textbf {\bibinfo {volume} {351}},\ \bibinfo {pages} {836} (\bibinfo {year} {2016})},\ \bibinfo {note} {\_eprint: https://www.science.org/doi/pdf/10.1126/science.aad8022}\BibitemShut {NoStop}%
\bibitem [{\citenamefont {Sushkov}\ \emph {et~al.}(2014{\natexlab{a}})\citenamefont {Sushkov}, \citenamefont {Chisholm}, \citenamefont {Lovchinsky}, \citenamefont {Kubo}, \citenamefont {Lo}, \citenamefont {Bennett}, \citenamefont {Hunger}, \citenamefont {Akimov}, \citenamefont {Walsworth}, \citenamefont {Park},\ and\ \citenamefont {Lukin}}]{sushkov_all-optical_2014}%
  \BibitemOpen
  \bibfield  {author} {\bibinfo {author} {\bibfnamefont {A.~O.}\ \bibnamefont {Sushkov}}, \bibinfo {author} {\bibfnamefont {N.}~\bibnamefont {Chisholm}}, \bibinfo {author} {\bibfnamefont {I.}~\bibnamefont {Lovchinsky}}, \bibinfo {author} {\bibfnamefont {M.}~\bibnamefont {Kubo}}, \bibinfo {author} {\bibfnamefont {P.~K.}\ \bibnamefont {Lo}}, \bibinfo {author} {\bibfnamefont {S.~D.}\ \bibnamefont {Bennett}}, \bibinfo {author} {\bibfnamefont {D.}~\bibnamefont {Hunger}}, \bibinfo {author} {\bibfnamefont {A.}~\bibnamefont {Akimov}}, \bibinfo {author} {\bibfnamefont {R.~L.}\ \bibnamefont {Walsworth}}, \bibinfo {author} {\bibfnamefont {H.}~\bibnamefont {Park}},\ and\ \bibinfo {author} {\bibfnamefont {M.~D.}\ \bibnamefont {Lukin}},\ }\bibfield  {title} {\bibinfo {title} {All-{Optical} {Sensing} of a {Single}-{Molecule} {Electron} {Spin}},\ }\href {https://doi.org/10.1021/nl502988n} {\bibfield  {journal} {\bibinfo  {journal} {Nano Letters}\ }\textbf {\bibinfo {volume} {14}},\ \bibinfo {pages} {6443} (\bibinfo {year}
  {2014}{\natexlab{a}})},\ \bibinfo {note} {publisher: American Chemical Society}\BibitemShut {NoStop}%
\bibitem [{\citenamefont {Shi}\ \emph {et~al.}(2018)\citenamefont {Shi}, \citenamefont {Kong}, \citenamefont {Zhao}, \citenamefont {Zhang}, \citenamefont {Chen}, \citenamefont {Chen}, \citenamefont {Zhang}, \citenamefont {Wang}, \citenamefont {Ye}, \citenamefont {Wang}, \citenamefont {Qin}, \citenamefont {Rong}, \citenamefont {Su}, \citenamefont {Wang}, \citenamefont {Qin},\ and\ \citenamefont {Du}}]{shi_single-dna_2018}%
  \BibitemOpen
  \bibfield  {author} {\bibinfo {author} {\bibfnamefont {F.}~\bibnamefont {Shi}}, \bibinfo {author} {\bibfnamefont {F.}~\bibnamefont {Kong}}, \bibinfo {author} {\bibfnamefont {P.}~\bibnamefont {Zhao}}, \bibinfo {author} {\bibfnamefont {X.}~\bibnamefont {Zhang}}, \bibinfo {author} {\bibfnamefont {M.}~\bibnamefont {Chen}}, \bibinfo {author} {\bibfnamefont {S.}~\bibnamefont {Chen}}, \bibinfo {author} {\bibfnamefont {Q.}~\bibnamefont {Zhang}}, \bibinfo {author} {\bibfnamefont {M.}~\bibnamefont {Wang}}, \bibinfo {author} {\bibfnamefont {X.}~\bibnamefont {Ye}}, \bibinfo {author} {\bibfnamefont {Z.}~\bibnamefont {Wang}}, \bibinfo {author} {\bibfnamefont {Z.}~\bibnamefont {Qin}}, \bibinfo {author} {\bibfnamefont {X.}~\bibnamefont {Rong}}, \bibinfo {author} {\bibfnamefont {J.}~\bibnamefont {Su}}, \bibinfo {author} {\bibfnamefont {P.}~\bibnamefont {Wang}}, \bibinfo {author} {\bibfnamefont {P.~Z.}\ \bibnamefont {Qin}},\ and\ \bibinfo {author} {\bibfnamefont {J.}~\bibnamefont {Du}},\ }\bibfield  {title} {\bibinfo {title}
  {Single-{DNA} electron spin resonance spectroscopy in aqueous solutions},\ }\href {https://doi.org/10.1038/s41592-018-0084-1} {\bibfield  {journal} {\bibinfo  {journal} {Nature Methods}\ }\textbf {\bibinfo {volume} {15}},\ \bibinfo {pages} {697} (\bibinfo {year} {2018})}\BibitemShut {NoStop}%
\bibitem [{\citenamefont {Tetienne}\ \emph {et~al.}(2014)\citenamefont {Tetienne}, \citenamefont {Hingant}, \citenamefont {Kim}, \citenamefont {Diez}, \citenamefont {Adam}, \citenamefont {Garcia}, \citenamefont {Roch}, \citenamefont {Rohart}, \citenamefont {Thiaville}, \citenamefont {Ravelosona},\ and\ \citenamefont {Jacques}}]{tetienne_nanoscale_2014}%
  \BibitemOpen
  \bibfield  {author} {\bibinfo {author} {\bibfnamefont {J.-P.}\ \bibnamefont {Tetienne}}, \bibinfo {author} {\bibfnamefont {T.}~\bibnamefont {Hingant}}, \bibinfo {author} {\bibfnamefont {J.-V.}\ \bibnamefont {Kim}}, \bibinfo {author} {\bibfnamefont {L.~H.}\ \bibnamefont {Diez}}, \bibinfo {author} {\bibfnamefont {J.-P.}\ \bibnamefont {Adam}}, \bibinfo {author} {\bibfnamefont {K.}~\bibnamefont {Garcia}}, \bibinfo {author} {\bibfnamefont {J.-F.}\ \bibnamefont {Roch}}, \bibinfo {author} {\bibfnamefont {S.}~\bibnamefont {Rohart}}, \bibinfo {author} {\bibfnamefont {A.}~\bibnamefont {Thiaville}}, \bibinfo {author} {\bibfnamefont {D.}~\bibnamefont {Ravelosona}},\ and\ \bibinfo {author} {\bibfnamefont {V.}~\bibnamefont {Jacques}},\ }\bibfield  {title} {\bibinfo {title} {Nanoscale imaging and control of domain-wall hopping with a nitrogen-vacancy center microscope},\ }\href {https://doi.org/10.1126/science.1250113} {\bibfield  {journal} {\bibinfo  {journal} {Science}\ }\textbf {\bibinfo {volume} {344}},\ \bibinfo
  {pages} {1366} (\bibinfo {year} {2014})},\ \bibinfo {note} {\_eprint: https://www.science.org/doi/pdf/10.1126/science.1250113}\BibitemShut {NoStop}%
\bibitem [{\citenamefont {Schlussel}\ \emph {et~al.}(2018)\citenamefont {Schlussel}, \citenamefont {Lenz}, \citenamefont {Rohner}, \citenamefont {Bar-Haim}, \citenamefont {Bougas}, \citenamefont {Groswasser}, \citenamefont {Kieschnick}, \citenamefont {Rozenberg}, \citenamefont {Thiel}, \citenamefont {Waxman}, \citenamefont {Meijer}, \citenamefont {Maletinsky}, \citenamefont {Budker},\ and\ \citenamefont {Folman}}]{schlussel_wide-field_2018}%
  \BibitemOpen
  \bibfield  {author} {\bibinfo {author} {\bibfnamefont {Y.}~\bibnamefont {Schlussel}}, \bibinfo {author} {\bibfnamefont {T.}~\bibnamefont {Lenz}}, \bibinfo {author} {\bibfnamefont {D.}~\bibnamefont {Rohner}}, \bibinfo {author} {\bibfnamefont {Y.}~\bibnamefont {Bar-Haim}}, \bibinfo {author} {\bibfnamefont {L.}~\bibnamefont {Bougas}}, \bibinfo {author} {\bibfnamefont {D.}~\bibnamefont {Groswasser}}, \bibinfo {author} {\bibfnamefont {M.}~\bibnamefont {Kieschnick}}, \bibinfo {author} {\bibfnamefont {E.}~\bibnamefont {Rozenberg}}, \bibinfo {author} {\bibfnamefont {L.}~\bibnamefont {Thiel}}, \bibinfo {author} {\bibfnamefont {A.}~\bibnamefont {Waxman}}, \bibinfo {author} {\bibfnamefont {J.}~\bibnamefont {Meijer}}, \bibinfo {author} {\bibfnamefont {P.}~\bibnamefont {Maletinsky}}, \bibinfo {author} {\bibfnamefont {D.}~\bibnamefont {Budker}},\ and\ \bibinfo {author} {\bibfnamefont {R.}~\bibnamefont {Folman}},\ }\bibfield  {title} {\bibinfo {title} {Wide-{Field} {Imaging} of {Superconductor} {Vortices} with {Electron}
  {Spins} in {Diamond}},\ }\href {https://doi.org/10.1103/PhysRevApplied.10.034032} {\bibfield  {journal} {\bibinfo  {journal} {Physical Review Applied}\ }\textbf {\bibinfo {volume} {10}},\ \bibinfo {pages} {034032} (\bibinfo {year} {2018})}\BibitemShut {NoStop}%
\bibitem [{\citenamefont {Bertelli}\ \emph {et~al.}(2020)\citenamefont {Bertelli}, \citenamefont {Carmiggelt}, \citenamefont {Yu}, \citenamefont {Simon}, \citenamefont {Pothoven}, \citenamefont {Bauer}, \citenamefont {Blanter}, \citenamefont {Aarts},\ and\ \citenamefont {Sar}}]{bertelli_magnetic_2020}%
  \BibitemOpen
  \bibfield  {author} {\bibinfo {author} {\bibfnamefont {I.}~\bibnamefont {Bertelli}}, \bibinfo {author} {\bibfnamefont {J.~J.}\ \bibnamefont {Carmiggelt}}, \bibinfo {author} {\bibfnamefont {T.}~\bibnamefont {Yu}}, \bibinfo {author} {\bibfnamefont {B.~G.}\ \bibnamefont {Simon}}, \bibinfo {author} {\bibfnamefont {C.~C.}\ \bibnamefont {Pothoven}}, \bibinfo {author} {\bibfnamefont {G.~E.~W.}\ \bibnamefont {Bauer}}, \bibinfo {author} {\bibfnamefont {Y.~M.}\ \bibnamefont {Blanter}}, \bibinfo {author} {\bibfnamefont {J.}~\bibnamefont {Aarts}},\ and\ \bibinfo {author} {\bibfnamefont {T.~v.~d.}\ \bibnamefont {Sar}},\ }\bibfield  {title} {\bibinfo {title} {Magnetic resonance imaging of spin-wave transport and interference in a magnetic insulator},\ }\href {https://doi.org/10.1126/sciadv.abd3556} {\bibfield  {journal} {\bibinfo  {journal} {Science Advances}\ }\textbf {\bibinfo {volume} {6}},\ \bibinfo {pages} {eabd3556} (\bibinfo {year} {2020})},\ \bibinfo {note} {\_eprint:
  https://www.science.org/doi/pdf/10.1126/sciadv.abd3556}\BibitemShut {NoStop}%
\bibitem [{\citenamefont {Degen}\ \emph {et~al.}(2009)\citenamefont {Degen}, \citenamefont {Poggio}, \citenamefont {Mamin}, \citenamefont {Rettner},\ and\ \citenamefont {Rugar}}]{degen_nanoscale_2009}%
  \BibitemOpen
  \bibfield  {author} {\bibinfo {author} {\bibfnamefont {C.~L.}\ \bibnamefont {Degen}}, \bibinfo {author} {\bibfnamefont {M.}~\bibnamefont {Poggio}}, \bibinfo {author} {\bibfnamefont {H.~J.}\ \bibnamefont {Mamin}}, \bibinfo {author} {\bibfnamefont {C.~T.}\ \bibnamefont {Rettner}},\ and\ \bibinfo {author} {\bibfnamefont {D.}~\bibnamefont {Rugar}},\ }\bibfield  {title} {\bibinfo {title} {Nanoscale magnetic resonance imaging},\ }\href {https://doi.org/10.1073/pnas.0812068106} {\bibfield  {journal} {\bibinfo  {journal} {Proceedings of the National Academy of Sciences}\ }\textbf {\bibinfo {volume} {106}},\ \bibinfo {pages} {1313} (\bibinfo {year} {2009})},\ \bibinfo {note} {\_eprint: https://www.pnas.org/doi/pdf/10.1073/pnas.0812068106}\BibitemShut {NoStop}%
\bibitem [{\citenamefont {Staudacher}\ \emph {et~al.}(2013)\citenamefont {Staudacher}, \citenamefont {Shi}, \citenamefont {Pezzagna}, \citenamefont {Meijer}, \citenamefont {Du}, \citenamefont {Meriles}, \citenamefont {Reinhard},\ and\ \citenamefont {Wrachtrup}}]{staudacher_nuclear_2013}%
  \BibitemOpen
  \bibfield  {author} {\bibinfo {author} {\bibfnamefont {T.}~\bibnamefont {Staudacher}}, \bibinfo {author} {\bibfnamefont {F.}~\bibnamefont {Shi}}, \bibinfo {author} {\bibfnamefont {S.}~\bibnamefont {Pezzagna}}, \bibinfo {author} {\bibfnamefont {J.}~\bibnamefont {Meijer}}, \bibinfo {author} {\bibfnamefont {J.}~\bibnamefont {Du}}, \bibinfo {author} {\bibfnamefont {C.~A.}\ \bibnamefont {Meriles}}, \bibinfo {author} {\bibfnamefont {F.}~\bibnamefont {Reinhard}},\ and\ \bibinfo {author} {\bibfnamefont {J.}~\bibnamefont {Wrachtrup}},\ }\bibfield  {title} {\bibinfo {title} {Nuclear {Magnetic} {Resonance} {Spectroscopy} on a (5-{Nanometer})$^{\textrm{3}}$ {Sample} {Volume}},\ }\href {https://doi.org/10.1126/science.1231675} {\bibfield  {journal} {\bibinfo  {journal} {Science}\ }\textbf {\bibinfo {volume} {339}},\ \bibinfo {pages} {561} (\bibinfo {year} {2013})},\ \bibinfo {note} {\_eprint: https://www.science.org/doi/pdf/10.1126/science.1231675}\BibitemShut {NoStop}%
\bibitem [{\citenamefont {Mamin}\ \emph {et~al.}(2013)\citenamefont {Mamin}, \citenamefont {Kim}, \citenamefont {Sherwood}, \citenamefont {Rettner}, \citenamefont {Ohno}, \citenamefont {Awschalom},\ and\ \citenamefont {Rugar}}]{mamin_nanoscale_2013}%
  \BibitemOpen
  \bibfield  {author} {\bibinfo {author} {\bibfnamefont {H.~J.}\ \bibnamefont {Mamin}}, \bibinfo {author} {\bibfnamefont {M.}~\bibnamefont {Kim}}, \bibinfo {author} {\bibfnamefont {M.~H.}\ \bibnamefont {Sherwood}}, \bibinfo {author} {\bibfnamefont {C.~T.}\ \bibnamefont {Rettner}}, \bibinfo {author} {\bibfnamefont {K.}~\bibnamefont {Ohno}}, \bibinfo {author} {\bibfnamefont {D.~D.}\ \bibnamefont {Awschalom}},\ and\ \bibinfo {author} {\bibfnamefont {D.}~\bibnamefont {Rugar}},\ }\bibfield  {title} {\bibinfo {title} {Nanoscale {Nuclear} {Magnetic} {Resonance} with a {Nitrogen}-{Vacancy} {Spin} {Sensor}},\ }\href {https://doi.org/10.1126/science.1231540} {\bibfield  {journal} {\bibinfo  {journal} {Science}\ }\textbf {\bibinfo {volume} {339}},\ \bibinfo {pages} {557} (\bibinfo {year} {2013})},\ \bibinfo {note} {\_eprint: https://www.science.org/doi/pdf/10.1126/science.1231540}\BibitemShut {NoStop}%
\bibitem [{\citenamefont {Myers}\ \emph {et~al.}(2014)\citenamefont {Myers}, \citenamefont {Das}, \citenamefont {Dartiailh}, \citenamefont {Ohno}, \citenamefont {Awschalom},\ and\ \citenamefont {Bleszynski~Jayich}}]{myers_probing_2014}%
  \BibitemOpen
  \bibfield  {author} {\bibinfo {author} {\bibfnamefont {B.}~\bibnamefont {Myers}}, \bibinfo {author} {\bibfnamefont {A.}~\bibnamefont {Das}}, \bibinfo {author} {\bibfnamefont {M.}~\bibnamefont {Dartiailh}}, \bibinfo {author} {\bibfnamefont {K.}~\bibnamefont {Ohno}}, \bibinfo {author} {\bibfnamefont {D.}~\bibnamefont {Awschalom}},\ and\ \bibinfo {author} {\bibfnamefont {A.}~\bibnamefont {Bleszynski~Jayich}},\ }\bibfield  {title} {\bibinfo {title} {Probing {Surface} {Noise} with {Depth}-{Calibrated} {Spins} in {Diamond}},\ }\href {https://doi.org/10.1103/PhysRevLett.113.027602} {\bibfield  {journal} {\bibinfo  {journal} {Physical Review Letters}\ }\textbf {\bibinfo {volume} {113}},\ \bibinfo {pages} {027602} (\bibinfo {year} {2014})}\BibitemShut {NoStop}%
\bibitem [{\citenamefont {Sangtawesin}\ \emph {et~al.}(2019)\citenamefont {Sangtawesin}, \citenamefont {Dwyer}, \citenamefont {Srinivasan}, \citenamefont {Allred}, \citenamefont {Rodgers}, \citenamefont {De~Greve}, \citenamefont {Stacey}, \citenamefont {Dontschuk}, \citenamefont {O’Donnell}, \citenamefont {Hu}, \citenamefont {Evans}, \citenamefont {Jaye}, \citenamefont {Fischer}, \citenamefont {Markham}, \citenamefont {Twitchen}, \citenamefont {Park}, \citenamefont {Lukin},\ and\ \citenamefont {De~Leon}}]{sangtawesin_origins_2019}%
  \BibitemOpen
  \bibfield  {author} {\bibinfo {author} {\bibfnamefont {S.}~\bibnamefont {Sangtawesin}}, \bibinfo {author} {\bibfnamefont {B.~L.}\ \bibnamefont {Dwyer}}, \bibinfo {author} {\bibfnamefont {S.}~\bibnamefont {Srinivasan}}, \bibinfo {author} {\bibfnamefont {J.~J.}\ \bibnamefont {Allred}}, \bibinfo {author} {\bibfnamefont {L.~V.}\ \bibnamefont {Rodgers}}, \bibinfo {author} {\bibfnamefont {K.}~\bibnamefont {De~Greve}}, \bibinfo {author} {\bibfnamefont {A.}~\bibnamefont {Stacey}}, \bibinfo {author} {\bibfnamefont {N.}~\bibnamefont {Dontschuk}}, \bibinfo {author} {\bibfnamefont {K.~M.}\ \bibnamefont {O’Donnell}}, \bibinfo {author} {\bibfnamefont {D.}~\bibnamefont {Hu}}, \bibinfo {author} {\bibfnamefont {D.~A.}\ \bibnamefont {Evans}}, \bibinfo {author} {\bibfnamefont {C.}~\bibnamefont {Jaye}}, \bibinfo {author} {\bibfnamefont {D.~A.}\ \bibnamefont {Fischer}}, \bibinfo {author} {\bibfnamefont {M.~L.}\ \bibnamefont {Markham}}, \bibinfo {author} {\bibfnamefont {D.~J.}\ \bibnamefont {Twitchen}}, \bibinfo {author}
  {\bibfnamefont {H.}~\bibnamefont {Park}}, \bibinfo {author} {\bibfnamefont {M.~D.}\ \bibnamefont {Lukin}},\ and\ \bibinfo {author} {\bibfnamefont {N.~P.}\ \bibnamefont {De~Leon}},\ }\bibfield  {title} {\bibinfo {title} {Origins of {Diamond} {Surface} {Noise} {Probed} by {Correlating} {Single}-{Spin} {Measurements} with {Surface} {Spectroscopy}},\ }\href {https://doi.org/10.1103/PhysRevX.9.031052} {\bibfield  {journal} {\bibinfo  {journal} {Physical Review X}\ }\textbf {\bibinfo {volume} {9}},\ \bibinfo {pages} {031052} (\bibinfo {year} {2019})}\BibitemShut {NoStop}%
\bibitem [{\citenamefont {Rondin}\ \emph {et~al.}(2010)\citenamefont {Rondin}, \citenamefont {Dantelle}, \citenamefont {Slablab}, \citenamefont {Grosshans}, \citenamefont {Treussart}, \citenamefont {Bergonzo}, \citenamefont {Perruchas}, \citenamefont {Gacoin}, \citenamefont {Chaigneau}, \citenamefont {Chang}, \citenamefont {Jacques},\ and\ \citenamefont {Roch}}]{rondin_surface-induced_2010}%
  \BibitemOpen
  \bibfield  {author} {\bibinfo {author} {\bibfnamefont {L.}~\bibnamefont {Rondin}}, \bibinfo {author} {\bibfnamefont {G.}~\bibnamefont {Dantelle}}, \bibinfo {author} {\bibfnamefont {A.}~\bibnamefont {Slablab}}, \bibinfo {author} {\bibfnamefont {F.}~\bibnamefont {Grosshans}}, \bibinfo {author} {\bibfnamefont {F.}~\bibnamefont {Treussart}}, \bibinfo {author} {\bibfnamefont {P.}~\bibnamefont {Bergonzo}}, \bibinfo {author} {\bibfnamefont {S.}~\bibnamefont {Perruchas}}, \bibinfo {author} {\bibfnamefont {T.}~\bibnamefont {Gacoin}}, \bibinfo {author} {\bibfnamefont {M.}~\bibnamefont {Chaigneau}}, \bibinfo {author} {\bibfnamefont {H.-C.}\ \bibnamefont {Chang}}, \bibinfo {author} {\bibfnamefont {V.}~\bibnamefont {Jacques}},\ and\ \bibinfo {author} {\bibfnamefont {J.-F.}\ \bibnamefont {Roch}},\ }\bibfield  {title} {\bibinfo {title} {Surface-induced charge state conversion of nitrogen-vacancy defects in nanodiamonds},\ }\href {https://doi.org/10.1103/PhysRevB.82.115449} {\bibfield  {journal} {\bibinfo  {journal}
  {Physical Review B}\ }\textbf {\bibinfo {volume} {82}},\ \bibinfo {pages} {115449} (\bibinfo {year} {2010})}\BibitemShut {NoStop}%
\bibitem [{\citenamefont {Yuan}\ \emph {et~al.}(2020)\citenamefont {Yuan}, \citenamefont {Fitzpatrick}, \citenamefont {Rodgers}, \citenamefont {Sangtawesin}, \citenamefont {Srinivasan},\ and\ \citenamefont {De~Leon}}]{yuan_charge_2020}%
  \BibitemOpen
  \bibfield  {author} {\bibinfo {author} {\bibfnamefont {Z.}~\bibnamefont {Yuan}}, \bibinfo {author} {\bibfnamefont {M.}~\bibnamefont {Fitzpatrick}}, \bibinfo {author} {\bibfnamefont {L.~V.~H.}\ \bibnamefont {Rodgers}}, \bibinfo {author} {\bibfnamefont {S.}~\bibnamefont {Sangtawesin}}, \bibinfo {author} {\bibfnamefont {S.}~\bibnamefont {Srinivasan}},\ and\ \bibinfo {author} {\bibfnamefont {N.~P.}\ \bibnamefont {De~Leon}},\ }\bibfield  {title} {\bibinfo {title} {Charge state dynamics and optically detected electron spin resonance contrast of shallow nitrogen-vacancy centers in diamond},\ }\href {https://doi.org/10.1103/PhysRevResearch.2.033263} {\bibfield  {journal} {\bibinfo  {journal} {Physical Review Research}\ }\textbf {\bibinfo {volume} {2}},\ \bibinfo {pages} {033263} (\bibinfo {year} {2020})}\BibitemShut {NoStop}%
\bibitem [{\citenamefont {Rosskopf}\ \emph {et~al.}(2014)\citenamefont {Rosskopf}, \citenamefont {Dussaux}, \citenamefont {Ohashi}, \citenamefont {Loretz}, \citenamefont {Schirhagl}, \citenamefont {Watanabe}, \citenamefont {Shikata}, \citenamefont {Itoh},\ and\ \citenamefont {Degen}}]{rosskopf_investigation_2014}%
  \BibitemOpen
  \bibfield  {author} {\bibinfo {author} {\bibfnamefont {T.}~\bibnamefont {Rosskopf}}, \bibinfo {author} {\bibfnamefont {A.}~\bibnamefont {Dussaux}}, \bibinfo {author} {\bibfnamefont {K.}~\bibnamefont {Ohashi}}, \bibinfo {author} {\bibfnamefont {M.}~\bibnamefont {Loretz}}, \bibinfo {author} {\bibfnamefont {R.}~\bibnamefont {Schirhagl}}, \bibinfo {author} {\bibfnamefont {H.}~\bibnamefont {Watanabe}}, \bibinfo {author} {\bibfnamefont {S.}~\bibnamefont {Shikata}}, \bibinfo {author} {\bibfnamefont {K.~M.}\ \bibnamefont {Itoh}},\ and\ \bibinfo {author} {\bibfnamefont {C.~L.}\ \bibnamefont {Degen}},\ }\bibfield  {title} {\bibinfo {title} {Investigation of {Surface} {Magnetic} {Noise} by {Shallow} {Spins} in {Diamond}},\ }\href {https://doi.org/10.1103/PhysRevLett.112.147602} {\bibfield  {journal} {\bibinfo  {journal} {Phys. Rev. Lett.}\ }\textbf {\bibinfo {volume} {112}},\ \bibinfo {pages} {147602} (\bibinfo {year} {2014})},\ \bibinfo {note} {publisher: American Physical Society}\BibitemShut {NoStop}%
\bibitem [{\citenamefont {Sushkov}\ \emph {et~al.}(2014{\natexlab{b}})\citenamefont {Sushkov}, \citenamefont {Lovchinsky}, \citenamefont {Chisholm}, \citenamefont {Walsworth}, \citenamefont {Park},\ and\ \citenamefont {Lukin}}]{sushkov_magnetic_2014}%
  \BibitemOpen
  \bibfield  {author} {\bibinfo {author} {\bibfnamefont {A.}~\bibnamefont {Sushkov}}, \bibinfo {author} {\bibfnamefont {I.}~\bibnamefont {Lovchinsky}}, \bibinfo {author} {\bibfnamefont {N.}~\bibnamefont {Chisholm}}, \bibinfo {author} {\bibfnamefont {R.}~\bibnamefont {Walsworth}}, \bibinfo {author} {\bibfnamefont {H.}~\bibnamefont {Park}},\ and\ \bibinfo {author} {\bibfnamefont {M.}~\bibnamefont {Lukin}},\ }\bibfield  {title} {\bibinfo {title} {Magnetic {Resonance} {Detection} of {Individual} {Proton} {Spins} {Using} {Quantum} {Reporters}},\ }\href {https://doi.org/10.1103/PhysRevLett.113.197601} {\bibfield  {journal} {\bibinfo  {journal} {Physical Review Letters}\ }\textbf {\bibinfo {volume} {113}},\ \bibinfo {pages} {197601} (\bibinfo {year} {2014}{\natexlab{b}})}\BibitemShut {NoStop}%
\bibitem [{\citenamefont {Zhang}\ \emph {et~al.}(2023)\citenamefont {Zhang}, \citenamefont {Joos}, \citenamefont {Bluvstein}, \citenamefont {Lyu},\ and\ \citenamefont {Bleszynski~Jayich}}]{zhang_reporter-spin-assisted_2023}%
  \BibitemOpen
  \bibfield  {author} {\bibinfo {author} {\bibfnamefont {Z.}~\bibnamefont {Zhang}}, \bibinfo {author} {\bibfnamefont {M.}~\bibnamefont {Joos}}, \bibinfo {author} {\bibfnamefont {D.}~\bibnamefont {Bluvstein}}, \bibinfo {author} {\bibfnamefont {Y.}~\bibnamefont {Lyu}},\ and\ \bibinfo {author} {\bibfnamefont {A.~C.}\ \bibnamefont {Bleszynski~Jayich}},\ }\bibfield  {title} {\bibinfo {title} {Reporter-{Spin}-{Assisted} {T} 1 {Relaxometry}},\ }\href {https://doi.org/10.1103/PhysRevApplied.19.L031004} {\bibfield  {journal} {\bibinfo  {journal} {Physical Review Applied}\ }\textbf {\bibinfo {volume} {19}},\ \bibinfo {pages} {L031004} (\bibinfo {year} {2023})}\BibitemShut {NoStop}%
\bibitem [{\citenamefont {Grotz}\ \emph {et~al.}(2011)\citenamefont {Grotz}, \citenamefont {Beck}, \citenamefont {Neumann}, \citenamefont {Naydenov}, \citenamefont {Reuter}, \citenamefont {Reinhard}, \citenamefont {Jelezko}, \citenamefont {Wrachtrup}, \citenamefont {Schweinfurth}, \citenamefont {Sarkar},\ and\ \citenamefont {Hemmer}}]{grotz_sensing_2011}%
  \BibitemOpen
  \bibfield  {author} {\bibinfo {author} {\bibfnamefont {B.}~\bibnamefont {Grotz}}, \bibinfo {author} {\bibfnamefont {J.}~\bibnamefont {Beck}}, \bibinfo {author} {\bibfnamefont {P.}~\bibnamefont {Neumann}}, \bibinfo {author} {\bibfnamefont {B.}~\bibnamefont {Naydenov}}, \bibinfo {author} {\bibfnamefont {R.}~\bibnamefont {Reuter}}, \bibinfo {author} {\bibfnamefont {F.}~\bibnamefont {Reinhard}}, \bibinfo {author} {\bibfnamefont {F.}~\bibnamefont {Jelezko}}, \bibinfo {author} {\bibfnamefont {J.}~\bibnamefont {Wrachtrup}}, \bibinfo {author} {\bibfnamefont {D.}~\bibnamefont {Schweinfurth}}, \bibinfo {author} {\bibfnamefont {B.}~\bibnamefont {Sarkar}},\ and\ \bibinfo {author} {\bibfnamefont {P.}~\bibnamefont {Hemmer}},\ }\bibfield  {title} {\bibinfo {title} {Sensing external spins with nitrogen-vacancy diamond},\ }\href {https://doi.org/10.1088/1367-2630/13/5/055004} {\bibfield  {journal} {\bibinfo  {journal} {New Journal of Physics}\ }\textbf {\bibinfo {volume} {13}},\ \bibinfo {pages} {055004} (\bibinfo {year}
  {2011})}\BibitemShut {NoStop}%
\bibitem [{\citenamefont {Mamin}\ \emph {et~al.}(2012)\citenamefont {Mamin}, \citenamefont {Sherwood},\ and\ \citenamefont {Rugar}}]{mamin_detecting_2012}%
  \BibitemOpen
  \bibfield  {author} {\bibinfo {author} {\bibfnamefont {H.~J.}\ \bibnamefont {Mamin}}, \bibinfo {author} {\bibfnamefont {M.~H.}\ \bibnamefont {Sherwood}},\ and\ \bibinfo {author} {\bibfnamefont {D.}~\bibnamefont {Rugar}},\ }\bibfield  {title} {\bibinfo {title} {Detecting external electron spins using nitrogen-vacancy centers},\ }\href {https://doi.org/10.1103/PhysRevB.86.195422} {\bibfield  {journal} {\bibinfo  {journal} {Physical Review B}\ }\textbf {\bibinfo {volume} {86}},\ \bibinfo {pages} {195422} (\bibinfo {year} {2012})}\BibitemShut {NoStop}%
\bibitem [{\citenamefont {Schaffry}\ \emph {et~al.}(2011)\citenamefont {Schaffry}, \citenamefont {Gauger}, \citenamefont {Morton},\ and\ \citenamefont {Benjamin}}]{schaffry_proposed_2011}%
  \BibitemOpen
  \bibfield  {author} {\bibinfo {author} {\bibfnamefont {M.}~\bibnamefont {Schaffry}}, \bibinfo {author} {\bibfnamefont {E.~M.}\ \bibnamefont {Gauger}}, \bibinfo {author} {\bibfnamefont {J.~J.~L.}\ \bibnamefont {Morton}},\ and\ \bibinfo {author} {\bibfnamefont {S.~C.}\ \bibnamefont {Benjamin}},\ }\bibfield  {title} {\bibinfo {title} {Proposed {Spin} {Amplification} for {Magnetic} {Sensors} {Employing} {Crystal} {Defects}},\ }\href {https://doi.org/10.1103/PhysRevLett.107.207210} {\bibfield  {journal} {\bibinfo  {journal} {Physical Review Letters}\ }\textbf {\bibinfo {volume} {107}},\ \bibinfo {pages} {207210} (\bibinfo {year} {2011})}\BibitemShut {NoStop}%
\bibitem [{\citenamefont {Dwyer}\ \emph {et~al.}(2022)\citenamefont {Dwyer}, \citenamefont {Rodgers}, \citenamefont {Urbach}, \citenamefont {Bluvstein}, \citenamefont {Sangtawesin}, \citenamefont {Zhou}, \citenamefont {Nassab}, \citenamefont {Fitzpatrick}, \citenamefont {Yuan}, \citenamefont {De~Greve}, \citenamefont {Peterson}, \citenamefont {Knowles}, \citenamefont {Sumarac}, \citenamefont {Chou}, \citenamefont {Gali}, \citenamefont {Dobrovitski}, \citenamefont {Lukin},\ and\ \citenamefont {De~Leon}}]{dwyer_probing_2022}%
  \BibitemOpen
  \bibfield  {author} {\bibinfo {author} {\bibfnamefont {B.~L.}\ \bibnamefont {Dwyer}}, \bibinfo {author} {\bibfnamefont {L.~V.}\ \bibnamefont {Rodgers}}, \bibinfo {author} {\bibfnamefont {E.~K.}\ \bibnamefont {Urbach}}, \bibinfo {author} {\bibfnamefont {D.}~\bibnamefont {Bluvstein}}, \bibinfo {author} {\bibfnamefont {S.}~\bibnamefont {Sangtawesin}}, \bibinfo {author} {\bibfnamefont {H.}~\bibnamefont {Zhou}}, \bibinfo {author} {\bibfnamefont {Y.}~\bibnamefont {Nassab}}, \bibinfo {author} {\bibfnamefont {M.}~\bibnamefont {Fitzpatrick}}, \bibinfo {author} {\bibfnamefont {Z.}~\bibnamefont {Yuan}}, \bibinfo {author} {\bibfnamefont {K.}~\bibnamefont {De~Greve}}, \bibinfo {author} {\bibfnamefont {E.~L.}\ \bibnamefont {Peterson}}, \bibinfo {author} {\bibfnamefont {H.}~\bibnamefont {Knowles}}, \bibinfo {author} {\bibfnamefont {T.}~\bibnamefont {Sumarac}}, \bibinfo {author} {\bibfnamefont {J.-P.}\ \bibnamefont {Chou}}, \bibinfo {author} {\bibfnamefont {A.}~\bibnamefont {Gali}}, \bibinfo {author} {\bibfnamefont
  {V.}~\bibnamefont {Dobrovitski}}, \bibinfo {author} {\bibfnamefont {M.~D.}\ \bibnamefont {Lukin}},\ and\ \bibinfo {author} {\bibfnamefont {N.~P.}\ \bibnamefont {De~Leon}},\ }\bibfield  {title} {\bibinfo {title} {Probing {Spin} {Dynamics} on {Diamond} {Surfaces} {Using} a {Single} {Quantum} {Sensor}},\ }\href {https://doi.org/10.1103/PRXQuantum.3.040328} {\bibfield  {journal} {\bibinfo  {journal} {PRX Quantum}\ }\textbf {\bibinfo {volume} {3}},\ \bibinfo {pages} {040328} (\bibinfo {year} {2022})}\BibitemShut {NoStop}%
\bibitem [{\citenamefont {Beaumont}\ \emph {et~al.}(1997)\citenamefont {Beaumont}, \citenamefont {Johnson},\ and\ \citenamefont {Parsons}}]{beaumont_excited_1997}%
  \BibitemOpen
  \bibfield  {author} {\bibinfo {author} {\bibfnamefont {P.~C.}\ \bibnamefont {Beaumont}}, \bibinfo {author} {\bibfnamefont {D.~G.}\ \bibnamefont {Johnson}},\ and\ \bibinfo {author} {\bibfnamefont {B.~J.}\ \bibnamefont {Parsons}},\ }\bibfield  {title} {\bibinfo {title} {Excited state and free radical properties of rhodamine dyes in aqueous solution: {A} laser flash photolysis and pulse radiolysis study},\ }\href {https://doi.org/https://doi.org/10.1016/S1010-6030(96)04591-1} {\bibfield  {journal} {\bibinfo  {journal} {Journal of Photochemistry and Photobiology A: Chemistry}\ }\textbf {\bibinfo {volume} {107}},\ \bibinfo {pages} {175} (\bibinfo {year} {1997})}\BibitemShut {NoStop}%
\bibitem [{\citenamefont {Zondervan}\ \emph {et~al.}(2004)\citenamefont {Zondervan}, \citenamefont {Kulzer}, \citenamefont {Kol'chenk},\ and\ \citenamefont {Orrit}}]{zondervan_photobleaching_2004}%
  \BibitemOpen
  \bibfield  {author} {\bibinfo {author} {\bibfnamefont {R.}~\bibnamefont {Zondervan}}, \bibinfo {author} {\bibfnamefont {F.}~\bibnamefont {Kulzer}}, \bibinfo {author} {\bibfnamefont {M.~A.}\ \bibnamefont {Kol'chenk}},\ and\ \bibinfo {author} {\bibfnamefont {M.}~\bibnamefont {Orrit}},\ }\bibfield  {title} {\bibinfo {title} {Photobleaching of {Rhodamine} {6G} in {Poly}(vinyl alcohol) at the {Ensemble} and {Single}-{Molecule} {Levels}},\ }\href {https://doi.org/10.1021/jp037222e} {\bibfield  {journal} {\bibinfo  {journal} {The Journal of Physical Chemistry A}\ }\textbf {\bibinfo {volume} {108}},\ \bibinfo {pages} {1657} (\bibinfo {year} {2004})},\ \bibinfo {note} {publisher: American Chemical Society}\BibitemShut {NoStop}%
\bibitem [{\citenamefont {van~de Linde}\ \emph {et~al.}(2011)\citenamefont {van~de Linde}, \citenamefont {Krstić}, \citenamefont {Prisner}, \citenamefont {Doose}, \citenamefont {Heilemann},\ and\ \citenamefont {Sauer}}]{van_de_linde_photoinduced_2011}%
  \BibitemOpen
  \bibfield  {author} {\bibinfo {author} {\bibfnamefont {S.}~\bibnamefont {van~de Linde}}, \bibinfo {author} {\bibfnamefont {I.}~\bibnamefont {Krstić}}, \bibinfo {author} {\bibfnamefont {T.}~\bibnamefont {Prisner}}, \bibinfo {author} {\bibfnamefont {S.}~\bibnamefont {Doose}}, \bibinfo {author} {\bibfnamefont {M.}~\bibnamefont {Heilemann}},\ and\ \bibinfo {author} {\bibfnamefont {M.}~\bibnamefont {Sauer}},\ }\bibfield  {title} {\bibinfo {title} {Photoinduced formation of reversible dye radicals and their impact on super-resolution imaging},\ }\href {https://doi.org/10.1039/C0PP00317D} {\bibfield  {journal} {\bibinfo  {journal} {Photochem. Photobiol. Sci.}\ }\textbf {\bibinfo {volume} {10}},\ \bibinfo {pages} {499} (\bibinfo {year} {2011})},\ \bibinfo {note} {publisher: The Royal Society of Chemistry}\BibitemShut {NoStop}%
\bibitem [{\citenamefont {Görner}(2008)}]{gorner_oxygen_2008}%
  \BibitemOpen
  \bibfield  {author} {\bibinfo {author} {\bibfnamefont {H.}~\bibnamefont {Görner}},\ }\bibfield  {title} {\bibinfo {title} {Oxygen uptake induced by electron transfer from donors to the triplet state of methylene blue and xanthene dyes in air-saturated aqueous solution},\ }\href {https://doi.org/10.1039/B712496A} {\bibfield  {journal} {\bibinfo  {journal} {Photochem. Photobiol. Sci.}\ }\textbf {\bibinfo {volume} {7}},\ \bibinfo {pages} {371} (\bibinfo {year} {2008})},\ \bibinfo {note} {publisher: The Royal Society of Chemistry}\BibitemShut {NoStop}%
\bibitem [{\citenamefont {Vogelsang}\ \emph {et~al.}(2008)\citenamefont {Vogelsang}, \citenamefont {Kasper}, \citenamefont {Steinhauer}, \citenamefont {Person}, \citenamefont {Heilemann}, \citenamefont {Sauer},\ and\ \citenamefont {Tinnefeld}}]{vogelsang_reducing_2008}%
  \BibitemOpen
  \bibfield  {author} {\bibinfo {author} {\bibfnamefont {J.}~\bibnamefont {Vogelsang}}, \bibinfo {author} {\bibfnamefont {R.}~\bibnamefont {Kasper}}, \bibinfo {author} {\bibfnamefont {C.}~\bibnamefont {Steinhauer}}, \bibinfo {author} {\bibfnamefont {B.}~\bibnamefont {Person}}, \bibinfo {author} {\bibfnamefont {M.}~\bibnamefont {Heilemann}}, \bibinfo {author} {\bibfnamefont {M.}~\bibnamefont {Sauer}},\ and\ \bibinfo {author} {\bibfnamefont {P.}~\bibnamefont {Tinnefeld}},\ }\bibfield  {title} {\bibinfo {title} {A {Reducing} and {Oxidizing} {System} {Minimizes} {Photobleaching} and {Blinking} of {Fluorescent} {Dyes}},\ }\href {https://doi.org/https://doi.org/10.1002/anie.200801518} {\bibfield  {journal} {\bibinfo  {journal} {Angewandte Chemie International Edition}\ }\textbf {\bibinfo {volume} {47}},\ \bibinfo {pages} {5465} (\bibinfo {year} {2008})}\BibitemShut {NoStop}%
\bibitem [{\citenamefont {Heilemann}\ \emph {et~al.}(2008)\citenamefont {Heilemann}, \citenamefont {Van De~Linde}, \citenamefont {Sch{\"u}ttpelz}, \citenamefont {Kasper}, \citenamefont {Seefeldt}, \citenamefont {Mukherjee}, \citenamefont {Tinnefeld},\ and\ \citenamefont {Sauer}}]{heilemann2008subdiffraction}%
  \BibitemOpen
  \bibfield  {author} {\bibinfo {author} {\bibfnamefont {M.}~\bibnamefont {Heilemann}}, \bibinfo {author} {\bibfnamefont {S.}~\bibnamefont {Van De~Linde}}, \bibinfo {author} {\bibfnamefont {M.}~\bibnamefont {Sch{\"u}ttpelz}}, \bibinfo {author} {\bibfnamefont {R.}~\bibnamefont {Kasper}}, \bibinfo {author} {\bibfnamefont {B.}~\bibnamefont {Seefeldt}}, \bibinfo {author} {\bibfnamefont {A.}~\bibnamefont {Mukherjee}}, \bibinfo {author} {\bibfnamefont {P.}~\bibnamefont {Tinnefeld}},\ and\ \bibinfo {author} {\bibfnamefont {M.}~\bibnamefont {Sauer}},\ }\bibfield  {title} {\bibinfo {title} {Subdiffraction-resolution fluorescence imaging with conventional fluorescent probes},\ }\href@noop {} {\bibfield  {journal} {\bibinfo  {journal} {Angewandte Chemie International Edition}\ }\textbf {\bibinfo {volume} {47}},\ \bibinfo {pages} {6172} (\bibinfo {year} {2008})}\BibitemShut {NoStop}%
\bibitem [{\citenamefont {Heilemann}\ \emph {et~al.}(2009)\citenamefont {Heilemann}, \citenamefont {Van De~Linde}, \citenamefont {Mukherjee},\ and\ \citenamefont {Sauer}}]{heilemann2009super}%
  \BibitemOpen
  \bibfield  {author} {\bibinfo {author} {\bibfnamefont {M.}~\bibnamefont {Heilemann}}, \bibinfo {author} {\bibfnamefont {S.}~\bibnamefont {Van De~Linde}}, \bibinfo {author} {\bibfnamefont {A.}~\bibnamefont {Mukherjee}},\ and\ \bibinfo {author} {\bibfnamefont {M.}~\bibnamefont {Sauer}},\ }\bibfield  {title} {\bibinfo {title} {Super-resolution imaging with small organic fluorophores},\ }\href@noop {} {\bibfield  {journal} {\bibinfo  {journal} {Angewandte Chemie International Edition}\ }\textbf {\bibinfo {volume} {48}},\ \bibinfo {pages} {6903} (\bibinfo {year} {2009})}\BibitemShut {NoStop}%
\bibitem [{\citenamefont {Zheng}\ \emph {et~al.}(2014)\citenamefont {Zheng}, \citenamefont {Juette}, \citenamefont {Jockusch}, \citenamefont {Wasserman}, \citenamefont {Zhou}, \citenamefont {Altman},\ and\ \citenamefont {Blanchard}}]{zheng2014ultra}%
  \BibitemOpen
  \bibfield  {author} {\bibinfo {author} {\bibfnamefont {Q.}~\bibnamefont {Zheng}}, \bibinfo {author} {\bibfnamefont {M.~F.}\ \bibnamefont {Juette}}, \bibinfo {author} {\bibfnamefont {S.}~\bibnamefont {Jockusch}}, \bibinfo {author} {\bibfnamefont {M.~R.}\ \bibnamefont {Wasserman}}, \bibinfo {author} {\bibfnamefont {Z.}~\bibnamefont {Zhou}}, \bibinfo {author} {\bibfnamefont {R.~B.}\ \bibnamefont {Altman}},\ and\ \bibinfo {author} {\bibfnamefont {S.~C.}\ \bibnamefont {Blanchard}},\ }\bibfield  {title} {\bibinfo {title} {Ultra-stable organic fluorophores for single-molecule research},\ }\href@noop {} {\bibfield  {journal} {\bibinfo  {journal} {Chemical Society Reviews}\ }\textbf {\bibinfo {volume} {43}},\ \bibinfo {pages} {1044} (\bibinfo {year} {2014})}\BibitemShut {NoStop}%
\bibitem [{\citenamefont {Brasselet}\ and\ \citenamefont {Lew}(2025)}]{brasselet2025single}%
  \BibitemOpen
  \bibfield  {author} {\bibinfo {author} {\bibfnamefont {S.}~\bibnamefont {Brasselet}}\ and\ \bibinfo {author} {\bibfnamefont {M.~D.}\ \bibnamefont {Lew}},\ }\bibfield  {title} {\bibinfo {title} {Single-molecule orientation and localization microscopy},\ }\href@noop {} {\bibfield  {journal} {\bibinfo  {journal} {Nature Photonics}\ ,\ \bibinfo {pages} {1}} (\bibinfo {year} {2025})}\BibitemShut {NoStop}%
\bibitem [{\citenamefont {Choi}\ \emph {et~al.}(2017)\citenamefont {Choi}, \citenamefont {Choi}, \citenamefont {Kucsko}, \citenamefont {Maurer}, \citenamefont {Shields}, \citenamefont {Sumiya}, \citenamefont {Onoda}, \citenamefont {Isoya}, \citenamefont {Demler}, \citenamefont {Jelezko}, \citenamefont {Yao},\ and\ \citenamefont {Lukin}}]{PhysRevLett.118.093601}%
  \BibitemOpen
  \bibfield  {author} {\bibinfo {author} {\bibfnamefont {J.}~\bibnamefont {Choi}}, \bibinfo {author} {\bibfnamefont {S.}~\bibnamefont {Choi}}, \bibinfo {author} {\bibfnamefont {G.}~\bibnamefont {Kucsko}}, \bibinfo {author} {\bibfnamefont {P.~C.}\ \bibnamefont {Maurer}}, \bibinfo {author} {\bibfnamefont {B.~J.}\ \bibnamefont {Shields}}, \bibinfo {author} {\bibfnamefont {H.}~\bibnamefont {Sumiya}}, \bibinfo {author} {\bibfnamefont {S.}~\bibnamefont {Onoda}}, \bibinfo {author} {\bibfnamefont {J.}~\bibnamefont {Isoya}}, \bibinfo {author} {\bibfnamefont {E.}~\bibnamefont {Demler}}, \bibinfo {author} {\bibfnamefont {F.}~\bibnamefont {Jelezko}}, \bibinfo {author} {\bibfnamefont {N.~Y.}\ \bibnamefont {Yao}},\ and\ \bibinfo {author} {\bibfnamefont {M.~D.}\ \bibnamefont {Lukin}},\ }\bibfield  {title} {\bibinfo {title} {Depolarization dynamics in a strongly interacting solid-state spin ensemble},\ }\href {https://doi.org/10.1103/PhysRevLett.118.093601} {\bibfield  {journal} {\bibinfo  {journal} {Phys. Rev. Lett.}\
  }\textbf {\bibinfo {volume} {118}},\ \bibinfo {pages} {093601} (\bibinfo {year} {2017})}\BibitemShut {NoStop}%
\bibitem [{\citenamefont {Sow}\ \emph {et~al.}(2020)\citenamefont {Sow}, \citenamefont {Steuer}, \citenamefont {Adekanye}, \citenamefont {Ginés}, \citenamefont {Mandal}, \citenamefont {Gilboa}, \citenamefont {Williams}, \citenamefont {Smith},\ and\ \citenamefont {Kapanidis}}]{sow_high-throughput_2020}%
  \BibitemOpen
  \bibfield  {author} {\bibinfo {author} {\bibfnamefont {M.}~\bibnamefont {Sow}}, \bibinfo {author} {\bibfnamefont {H.}~\bibnamefont {Steuer}}, \bibinfo {author} {\bibfnamefont {S.}~\bibnamefont {Adekanye}}, \bibinfo {author} {\bibfnamefont {L.}~\bibnamefont {Ginés}}, \bibinfo {author} {\bibfnamefont {S.}~\bibnamefont {Mandal}}, \bibinfo {author} {\bibfnamefont {B.}~\bibnamefont {Gilboa}}, \bibinfo {author} {\bibfnamefont {O.~A.}\ \bibnamefont {Williams}}, \bibinfo {author} {\bibfnamefont {J.~M.}\ \bibnamefont {Smith}},\ and\ \bibinfo {author} {\bibfnamefont {A.~N.}\ \bibnamefont {Kapanidis}},\ }\bibfield  {title} {\bibinfo {title} {High-throughput nitrogen-vacancy center imaging for nanodiamond photophysical characterization and {pH} nanosensing},\ }\href {https://doi.org/10.1039/D0NR05931E} {\bibfield  {journal} {\bibinfo  {journal} {Nanoscale}\ }\textbf {\bibinfo {volume} {12}},\ \bibinfo {pages} {21821} (\bibinfo {year} {2020})},\ \bibinfo {note} {publisher: The Royal Society of Chemistry}\BibitemShut
  {NoStop}%
\bibitem [{\citenamefont {Zhang}\ \emph {et~al.}(2021)\citenamefont {Zhang}, \citenamefont {Ghosh}, \citenamefont {Saxena},\ and\ \citenamefont {Dutt}}]{zhang_nanoscale_2021}%
  \BibitemOpen
  \bibfield  {author} {\bibinfo {author} {\bibfnamefont {K.}~\bibnamefont {Zhang}}, \bibinfo {author} {\bibfnamefont {S.}~\bibnamefont {Ghosh}}, \bibinfo {author} {\bibfnamefont {S.}~\bibnamefont {Saxena}},\ and\ \bibinfo {author} {\bibfnamefont {M.~V.~G.}\ \bibnamefont {Dutt}},\ }\bibfield  {title} {\bibinfo {title} {Nanoscale spin detection of copper ions using double electron-electron resonance at room temperature},\ }\href {https://doi.org/10.1103/PhysRevB.104.224412} {\bibfield  {journal} {\bibinfo  {journal} {Physical Review B}\ }\textbf {\bibinfo {volume} {104}},\ \bibinfo {pages} {224412} (\bibinfo {year} {2021})}\BibitemShut {NoStop}%
\bibitem [{\citenamefont {Bluvstein}\ \emph {et~al.}(2019)\citenamefont {Bluvstein}, \citenamefont {Zhang}, \citenamefont {McLellan}, \citenamefont {Williams},\ and\ \citenamefont {Jayich}}]{bluvstein_extending_2019}%
  \BibitemOpen
  \bibfield  {author} {\bibinfo {author} {\bibfnamefont {D.}~\bibnamefont {Bluvstein}}, \bibinfo {author} {\bibfnamefont {Z.}~\bibnamefont {Zhang}}, \bibinfo {author} {\bibfnamefont {C.~A.}\ \bibnamefont {McLellan}}, \bibinfo {author} {\bibfnamefont {N.~R.}\ \bibnamefont {Williams}},\ and\ \bibinfo {author} {\bibfnamefont {A.~C.~B.}\ \bibnamefont {Jayich}},\ }\bibfield  {title} {\bibinfo {title} {Extending the {Quantum} {Coherence} of a {Near}-{Surface} {Qubit} by {Coherently} {Driving} the {Paramagnetic} {Surface} {Environment}},\ }\href {https://doi.org/10.1103/PhysRevLett.123.146804} {\bibfield  {journal} {\bibinfo  {journal} {Physical Review Letters}\ }\textbf {\bibinfo {volume} {123}},\ \bibinfo {pages} {146804} (\bibinfo {year} {2019})}\BibitemShut {NoStop}%
\bibitem [{\citenamefont {Li}\ \emph {et~al.}(2021)\citenamefont {Li}, \citenamefont {Zheng}, \citenamefont {Peng}, \citenamefont {Kamiya}, \citenamefont {Niki}, \citenamefont {Stepanov}, \citenamefont {Jarmola}, \citenamefont {Shimizu}, \citenamefont {Takahashi}, \citenamefont {Wickenbrock},\ and\ \citenamefont {Budker}}]{li_determination_2021}%
  \BibitemOpen
  \bibfield  {author} {\bibinfo {author} {\bibfnamefont {S.}~\bibnamefont {Li}}, \bibinfo {author} {\bibfnamefont {H.}~\bibnamefont {Zheng}}, \bibinfo {author} {\bibfnamefont {Z.}~\bibnamefont {Peng}}, \bibinfo {author} {\bibfnamefont {M.}~\bibnamefont {Kamiya}}, \bibinfo {author} {\bibfnamefont {T.}~\bibnamefont {Niki}}, \bibinfo {author} {\bibfnamefont {V.}~\bibnamefont {Stepanov}}, \bibinfo {author} {\bibfnamefont {A.}~\bibnamefont {Jarmola}}, \bibinfo {author} {\bibfnamefont {Y.}~\bibnamefont {Shimizu}}, \bibinfo {author} {\bibfnamefont {S.}~\bibnamefont {Takahashi}}, \bibinfo {author} {\bibfnamefont {A.}~\bibnamefont {Wickenbrock}},\ and\ \bibinfo {author} {\bibfnamefont {D.}~\bibnamefont {Budker}},\ }\bibfield  {title} {\bibinfo {title} {Determination of local defect density in diamond by double electron-electron resonance},\ }\href {https://doi.org/10.1103/PhysRevB.104.094307} {\bibfield  {journal} {\bibinfo  {journal} {Physical Review B}\ }\textbf {\bibinfo {volume} {104}},\ \bibinfo {pages} {094307}
  (\bibinfo {year} {2021})}\BibitemShut {NoStop}%
\bibitem [{\citenamefont {Yu}\ \emph {et~al.}(2025)\citenamefont {Yu}, \citenamefont {Villafranca}, \citenamefont {Wang}, \citenamefont {Jones}, \citenamefont {Xie}, \citenamefont {Nagura}, \citenamefont {Chi-Durán}, \citenamefont {Delegan}, \citenamefont {Martinson}, \citenamefont {Flatté}, \citenamefont {Candido}, \citenamefont {Galli},\ and\ \citenamefont {Maurer}}]{yu_engineering_2025}%
  \BibitemOpen
  \bibfield  {author} {\bibinfo {author} {\bibfnamefont {X.}~\bibnamefont {Yu}}, \bibinfo {author} {\bibfnamefont {E.~J.}\ \bibnamefont {Villafranca}}, \bibinfo {author} {\bibfnamefont {S.}~\bibnamefont {Wang}}, \bibinfo {author} {\bibfnamefont {J.~C.}\ \bibnamefont {Jones}}, \bibinfo {author} {\bibfnamefont {M.}~\bibnamefont {Xie}}, \bibinfo {author} {\bibfnamefont {J.}~\bibnamefont {Nagura}}, \bibinfo {author} {\bibfnamefont {I.}~\bibnamefont {Chi-Durán}}, \bibinfo {author} {\bibfnamefont {N.}~\bibnamefont {Delegan}}, \bibinfo {author} {\bibfnamefont {A.~B.~F.}\ \bibnamefont {Martinson}}, \bibinfo {author} {\bibfnamefont {M.~E.}\ \bibnamefont {Flatté}}, \bibinfo {author} {\bibfnamefont {D.~R.}\ \bibnamefont {Candido}}, \bibinfo {author} {\bibfnamefont {G.}~\bibnamefont {Galli}},\ and\ \bibinfo {author} {\bibfnamefont {P.~C.}\ \bibnamefont {Maurer}},\ }\href {https://doi.org/10.48550/arXiv.2504.08883} {\bibinfo {title} {Engineering {Dark} {Spin}-{Free} {Diamond} {Interfaces}}} (\bibinfo {year} {2025}),\
  \bibinfo {note} {arXiv:2504.08883 [quant-ph]}\BibitemShut {NoStop}%
\bibitem [{\citenamefont {Espinós}\ \emph {et~al.}(2024)\citenamefont {Espinós}, \citenamefont {Munuera-Javaloy}, \citenamefont {Panadero}, \citenamefont {Acedo}, \citenamefont {Puebla}, \citenamefont {Casanova},\ and\ \citenamefont {Torrontegui}}]{espinos_enhancing_2024}%
  \BibitemOpen
  \bibfield  {author} {\bibinfo {author} {\bibfnamefont {H.}~\bibnamefont {Espinós}}, \bibinfo {author} {\bibfnamefont {C.}~\bibnamefont {Munuera-Javaloy}}, \bibinfo {author} {\bibfnamefont {I.}~\bibnamefont {Panadero}}, \bibinfo {author} {\bibfnamefont {P.}~\bibnamefont {Acedo}}, \bibinfo {author} {\bibfnamefont {R.}~\bibnamefont {Puebla}}, \bibinfo {author} {\bibfnamefont {J.}~\bibnamefont {Casanova}},\ and\ \bibinfo {author} {\bibfnamefont {E.}~\bibnamefont {Torrontegui}},\ }\bibfield  {title} {\bibinfo {title} {Enhancing polarization transfer from nitrogen-vacancy centers to external nuclear spins via dangling bond mediators},\ }\href {https://doi.org/10.1038/s42005-024-01536-6} {\bibfield  {journal} {\bibinfo  {journal} {Communications Physics}\ }\textbf {\bibinfo {volume} {7}},\ \bibinfo {pages} {42} (\bibinfo {year} {2024})}\BibitemShut {NoStop}%
\bibitem [{\citenamefont {Belthangady}\ \emph {et~al.}(2013)\citenamefont {Belthangady}, \citenamefont {Bar-Gill}, \citenamefont {Pham}, \citenamefont {Arai}, \citenamefont {Le~Sage}, \citenamefont {Cappellaro},\ and\ \citenamefont {Walsworth}}]{belthangady_dressed-state_2013}%
  \BibitemOpen
  \bibfield  {author} {\bibinfo {author} {\bibfnamefont {C.}~\bibnamefont {Belthangady}}, \bibinfo {author} {\bibfnamefont {N.}~\bibnamefont {Bar-Gill}}, \bibinfo {author} {\bibfnamefont {L.~M.}\ \bibnamefont {Pham}}, \bibinfo {author} {\bibfnamefont {K.}~\bibnamefont {Arai}}, \bibinfo {author} {\bibfnamefont {D.}~\bibnamefont {Le~Sage}}, \bibinfo {author} {\bibfnamefont {P.}~\bibnamefont {Cappellaro}},\ and\ \bibinfo {author} {\bibfnamefont {R.~L.}\ \bibnamefont {Walsworth}},\ }\bibfield  {title} {\bibinfo {title} {Dressed-{State} {Resonant} {Coupling} between {Bright} and {Dark} {Spins} in {Diamond}},\ }\href {https://doi.org/10.1103/PhysRevLett.110.157601} {\bibfield  {journal} {\bibinfo  {journal} {Physical Review Letters}\ }\textbf {\bibinfo {volume} {110}},\ \bibinfo {pages} {157601} (\bibinfo {year} {2013})}\BibitemShut {NoStop}%
\bibitem [{\citenamefont {Dahlberg}\ and\ \citenamefont {Moerner}(2021)}]{dahlberg2021cryogenic}%
  \BibitemOpen
  \bibfield  {author} {\bibinfo {author} {\bibfnamefont {P.~D.}\ \bibnamefont {Dahlberg}}\ and\ \bibinfo {author} {\bibfnamefont {W.}~\bibnamefont {Moerner}},\ }\bibfield  {title} {\bibinfo {title} {Cryogenic super-resolution fluorescence and electron microscopy correlated at the nanoscale},\ }\href@noop {} {\bibfield  {journal} {\bibinfo  {journal} {Annual Review of Physical Chemistry}\ }\textbf {\bibinfo {volume} {72}},\ \bibinfo {pages} {253} (\bibinfo {year} {2021})}\BibitemShut {NoStop}%
\bibitem [{\citenamefont {Davis}\ \emph {et~al.}(2023)\citenamefont {Davis}, \citenamefont {Ye}, \citenamefont {Machado}, \citenamefont {Meynell}, \citenamefont {Wu}, \citenamefont {Mittiga}, \citenamefont {Schenken}, \citenamefont {Joos}, \citenamefont {Kobrin}, \citenamefont {Lyu} \emph {et~al.}}]{davis2023probing}%
  \BibitemOpen
  \bibfield  {author} {\bibinfo {author} {\bibfnamefont {E.~J.}\ \bibnamefont {Davis}}, \bibinfo {author} {\bibfnamefont {B.}~\bibnamefont {Ye}}, \bibinfo {author} {\bibfnamefont {F.}~\bibnamefont {Machado}}, \bibinfo {author} {\bibfnamefont {S.~A.}\ \bibnamefont {Meynell}}, \bibinfo {author} {\bibfnamefont {W.}~\bibnamefont {Wu}}, \bibinfo {author} {\bibfnamefont {T.}~\bibnamefont {Mittiga}}, \bibinfo {author} {\bibfnamefont {W.}~\bibnamefont {Schenken}}, \bibinfo {author} {\bibfnamefont {M.}~\bibnamefont {Joos}}, \bibinfo {author} {\bibfnamefont {B.}~\bibnamefont {Kobrin}}, \bibinfo {author} {\bibfnamefont {Y.}~\bibnamefont {Lyu}}, \emph {et~al.},\ }\bibfield  {title} {\bibinfo {title} {Probing many-body dynamics in a two-dimensional dipolar spin ensemble},\ }\href@noop {} {\bibfield  {journal} {\bibinfo  {journal} {Nature Physics}\ }\textbf {\bibinfo {volume} {19}},\ \bibinfo {pages} {836} (\bibinfo {year} {2023})}\BibitemShut {NoStop}%
\bibitem [{\citenamefont {Hughes}\ \emph {et~al.}(2025)\citenamefont {Hughes}, \citenamefont {Meynell}, \citenamefont {Wu}, \citenamefont {Parthasarathy}, \citenamefont {Chen}, \citenamefont {Zhang}, \citenamefont {Wang}, \citenamefont {Davis}, \citenamefont {Mukherjee}, \citenamefont {Yao},\ and\ \citenamefont {Jayich}}]{HughesPRX2025}%
  \BibitemOpen
  \bibfield  {author} {\bibinfo {author} {\bibfnamefont {L.~B.}\ \bibnamefont {Hughes}}, \bibinfo {author} {\bibfnamefont {S.~A.}\ \bibnamefont {Meynell}}, \bibinfo {author} {\bibfnamefont {W.}~\bibnamefont {Wu}}, \bibinfo {author} {\bibfnamefont {S.}~\bibnamefont {Parthasarathy}}, \bibinfo {author} {\bibfnamefont {L.}~\bibnamefont {Chen}}, \bibinfo {author} {\bibfnamefont {Z.}~\bibnamefont {Zhang}}, \bibinfo {author} {\bibfnamefont {Z.}~\bibnamefont {Wang}}, \bibinfo {author} {\bibfnamefont {E.~J.}\ \bibnamefont {Davis}}, \bibinfo {author} {\bibfnamefont {K.}~\bibnamefont {Mukherjee}}, \bibinfo {author} {\bibfnamefont {N.~Y.}\ \bibnamefont {Yao}},\ and\ \bibinfo {author} {\bibfnamefont {A.~C.~B.}\ \bibnamefont {Jayich}},\ }\bibfield  {title} {\bibinfo {title} {Strongly interacting, two-dimensional, dipolar spin ensembles in (111)-oriented diamond},\ }\href {https://doi.org/10.1103/PhysRevX.15.021035} {\bibfield  {journal} {\bibinfo  {journal} {Phys. Rev. X}\ }\textbf {\bibinfo {volume} {15}},\ \bibinfo
  {pages} {021035} (\bibinfo {year} {2025})}\BibitemShut {NoStop}%
\bibitem [{\citenamefont {Hell}\ and\ \citenamefont {Wichmann}(1994)}]{hell1994breaking}%
  \BibitemOpen
  \bibfield  {author} {\bibinfo {author} {\bibfnamefont {S.~W.}\ \bibnamefont {Hell}}\ and\ \bibinfo {author} {\bibfnamefont {J.}~\bibnamefont {Wichmann}},\ }\bibfield  {title} {\bibinfo {title} {Breaking the diffraction resolution limit by stimulated emission: stimulated-emission-depletion fluorescence microscopy},\ }\href@noop {} {\bibfield  {journal} {\bibinfo  {journal} {Optics letters}\ }\textbf {\bibinfo {volume} {19}},\ \bibinfo {pages} {780} (\bibinfo {year} {1994})}\BibitemShut {NoStop}%
\bibitem [{\citenamefont {Schmied}\ \emph {et~al.}(2012)\citenamefont {Schmied}, \citenamefont {Gietl}, \citenamefont {Holzmeister}, \citenamefont {Forthmann}, \citenamefont {Steinhauer}, \citenamefont {Dammeyer},\ and\ \citenamefont {Tinnefeld}}]{schmied2012fluorescence}%
  \BibitemOpen
  \bibfield  {author} {\bibinfo {author} {\bibfnamefont {J.~J.}\ \bibnamefont {Schmied}}, \bibinfo {author} {\bibfnamefont {A.}~\bibnamefont {Gietl}}, \bibinfo {author} {\bibfnamefont {P.}~\bibnamefont {Holzmeister}}, \bibinfo {author} {\bibfnamefont {C.}~\bibnamefont {Forthmann}}, \bibinfo {author} {\bibfnamefont {C.}~\bibnamefont {Steinhauer}}, \bibinfo {author} {\bibfnamefont {T.}~\bibnamefont {Dammeyer}},\ and\ \bibinfo {author} {\bibfnamefont {P.}~\bibnamefont {Tinnefeld}},\ }\bibfield  {title} {\bibinfo {title} {Fluorescence and super-resolution standards based on dna origami},\ }\href@noop {} {\bibfield  {journal} {\bibinfo  {journal} {Nature methods}\ }\textbf {\bibinfo {volume} {9}},\ \bibinfo {pages} {1133} (\bibinfo {year} {2012})}\BibitemShut {NoStop}%
\bibitem [{\citenamefont {Bayliss}\ \emph {et~al.}(2020)\citenamefont {Bayliss}, \citenamefont {Laorenza}, \citenamefont {Mintun}, \citenamefont {Kovos}, \citenamefont {Freedman},\ and\ \citenamefont {Awschalom}}]{bayliss_laorenza_mintun_kovos_freedman_awschalom_2020}%
  \BibitemOpen
  \bibfield  {author} {\bibinfo {author} {\bibfnamefont {S.~L.}\ \bibnamefont {Bayliss}}, \bibinfo {author} {\bibfnamefont {D.~W.}\ \bibnamefont {Laorenza}}, \bibinfo {author} {\bibfnamefont {P.~J.}\ \bibnamefont {Mintun}}, \bibinfo {author} {\bibfnamefont {B.~D.}\ \bibnamefont {Kovos}}, \bibinfo {author} {\bibfnamefont {D.~E.}\ \bibnamefont {Freedman}},\ and\ \bibinfo {author} {\bibfnamefont {D.~D.}\ \bibnamefont {Awschalom}},\ }\bibfield  {title} {\bibinfo {title} {Optically addressable molecular spins for quantum information processing},\ }\href {https://doi.org/https://doi.org/10.1126/science.abb9352} {\bibfield  {journal} {\bibinfo  {journal} {Science}\ ,\ \bibinfo {pages} {eabb9352}} (\bibinfo {year} {2020})}\BibitemShut {NoStop}%
\bibitem [{\citenamefont {Kopp}\ \emph {et~al.}(2024)\citenamefont {Kopp}, \citenamefont {Nakamura}, \citenamefont {Phelan}, \citenamefont {Poh}, \citenamefont {Tyndall}, \citenamefont {Brown}, \citenamefont {Huang}, \citenamefont {Yuen-Zhou}, \citenamefont {Krzyaniak},\ and\ \citenamefont {Wasielewski}}]{kopp_nakamura_phelan_poh_tyndall_brown_huang_yuen-zhou_krzyaniak_wasielewski_2024}%
  \BibitemOpen
  \bibfield  {author} {\bibinfo {author} {\bibfnamefont {S.~M.}\ \bibnamefont {Kopp}}, \bibinfo {author} {\bibfnamefont {S.}~\bibnamefont {Nakamura}}, \bibinfo {author} {\bibfnamefont {B.~T.}\ \bibnamefont {Phelan}}, \bibinfo {author} {\bibfnamefont {Y.~R.}\ \bibnamefont {Poh}}, \bibinfo {author} {\bibfnamefont {S.~B.}\ \bibnamefont {Tyndall}}, \bibinfo {author} {\bibfnamefont {P.~J.}\ \bibnamefont {Brown}}, \bibinfo {author} {\bibfnamefont {Y.}~\bibnamefont {Huang}}, \bibinfo {author} {\bibfnamefont {J.}~\bibnamefont {Yuen-Zhou}}, \bibinfo {author} {\bibfnamefont {M.~D.}\ \bibnamefont {Krzyaniak}},\ and\ \bibinfo {author} {\bibfnamefont {M.~R.}\ \bibnamefont {Wasielewski}},\ }\bibfield  {title} {\bibinfo {title} {Luminescent organic triplet diradicals as optically addressable molecular qubits},\ }\href@noop {} {\bibfield  {journal} {\bibinfo  {journal} {Journal of the American Chemical Society}\ }\textbf {\bibinfo {volume} {146}} (\bibinfo {year} {2024})}\BibitemShut {NoStop}%
\bibitem [{\citenamefont {Chowdhury}\ \emph {et~al.}(2024)\citenamefont {Chowdhury}, \citenamefont {Murto}, \citenamefont {Panjwani}, \citenamefont {Sun}, \citenamefont {Ghosh}, \citenamefont {Boeije}, \citenamefont {Derkach}, \citenamefont {Woo}, \citenamefont {Millington}, \citenamefont {Congrave}, \citenamefont {Fu}, \citenamefont {Mustafa}, \citenamefont {Monteverde}, \citenamefont {Cerdá}, \citenamefont {Behrends}, \citenamefont {Rao}, \citenamefont {Beljonne}, \citenamefont {Chepelianskii}, \citenamefont {Bronstein},\ and\ \citenamefont {Friend}}]{chowdhury_murto_panjwani_sun_ghosh_boeije_derkach_woo_millington_congrave_etal._2024}%
  \BibitemOpen
  \bibfield  {author} {\bibinfo {author} {\bibfnamefont {R.}~\bibnamefont {Chowdhury}}, \bibinfo {author} {\bibfnamefont {P.}~\bibnamefont {Murto}}, \bibinfo {author} {\bibfnamefont {N.~A.}\ \bibnamefont {Panjwani}}, \bibinfo {author} {\bibfnamefont {Y.}~\bibnamefont {Sun}}, \bibinfo {author} {\bibfnamefont {P.}~\bibnamefont {Ghosh}}, \bibinfo {author} {\bibfnamefont {Y.}~\bibnamefont {Boeije}}, \bibinfo {author} {\bibfnamefont {V.}~\bibnamefont {Derkach}}, \bibinfo {author} {\bibfnamefont {S.-J.}\ \bibnamefont {Woo}}, \bibinfo {author} {\bibfnamefont {O.}~\bibnamefont {Millington}}, \bibinfo {author} {\bibfnamefont {D.~G.}\ \bibnamefont {Congrave}}, \bibinfo {author} {\bibfnamefont {Y.}~\bibnamefont {Fu}}, \bibinfo {author} {\bibfnamefont {T.~B.~E.}\ \bibnamefont {Mustafa}}, \bibinfo {author} {\bibfnamefont {M.}~\bibnamefont {Monteverde}}, \bibinfo {author} {\bibfnamefont {J.}~\bibnamefont {Cerdá}}, \bibinfo {author} {\bibfnamefont {J.}~\bibnamefont {Behrends}}, \bibinfo {author} {\bibfnamefont
  {A.}~\bibnamefont {Rao}}, \bibinfo {author} {\bibfnamefont {D.}~\bibnamefont {Beljonne}}, \bibinfo {author} {\bibfnamefont {A.}~\bibnamefont {Chepelianskii}}, \bibinfo {author} {\bibfnamefont {H.}~\bibnamefont {Bronstein}},\ and\ \bibinfo {author} {\bibfnamefont {R.~H.}\ \bibnamefont {Friend}},\ }\href {https://arxiv.org/abs/2406.03365} {\bibinfo {title} {Optical read and write of spin states in organic diradicals}} (\bibinfo {year} {2024})\BibitemShut {NoStop}%
\bibitem [{\citenamefont {Sutcliffe}\ \emph {et~al.}(2024)\citenamefont {Sutcliffe}, \citenamefont {Kazmierczak},\ and\ \citenamefont {Hadt}}]{sutcliffe_kazmierczak_hadt_2024}%
  \BibitemOpen
  \bibfield  {author} {\bibinfo {author} {\bibfnamefont {E.}~\bibnamefont {Sutcliffe}}, \bibinfo {author} {\bibfnamefont {N.~P.}\ \bibnamefont {Kazmierczak}},\ and\ \bibinfo {author} {\bibfnamefont {R.~G.}\ \bibnamefont {Hadt}},\ }\bibfield  {title} {\bibinfo {title} {Ultrafast all-optical coherence of molecular electron spins in room-temperature water solution},\ }\href {https://doi.org/https://doi.org/10.1126/science.ads0512} {\bibfield  {journal} {\bibinfo  {journal} {Science}\ }\textbf {\bibinfo {volume} {386}},\ \bibinfo {pages} {888–892} (\bibinfo {year} {2024})}\BibitemShut {NoStop}%
\bibitem [{\citenamefont {Zhou}\ \emph {et~al.}(2024)\citenamefont {Zhou}, \citenamefont {Sun},\ and\ \citenamefont {Sun}}]{zhou_sun_sun_2024}%
  \BibitemOpen
  \bibfield  {author} {\bibinfo {author} {\bibfnamefont {A.}~\bibnamefont {Zhou}}, \bibinfo {author} {\bibfnamefont {Z.}~\bibnamefont {Sun}},\ and\ \bibinfo {author} {\bibfnamefont {L.}~\bibnamefont {Sun}},\ }\bibfield  {title} {\bibinfo {title} {Stable organic radical qubits and their applications in quantum information science},\ }\href {https://doi.org/https://doi.org/10.1016/j.xinn.2024.100662} {\bibfield  {journal} {\bibinfo  {journal} {The Innovation}\ }\textbf {\bibinfo {volume} {5}},\ \bibinfo {pages} {100662–100662} (\bibinfo {year} {2024})}\BibitemShut {NoStop}%
\bibitem [{\citenamefont {Feder}\ \emph {et~al.}(2025)\citenamefont {Feder}, \citenamefont {Soloway}, \citenamefont {Verma}, \citenamefont {Geng}, \citenamefont {Wang}, \citenamefont {Kifle}, \citenamefont {Riendeau}, \citenamefont {Tsaturyan}, \citenamefont {Weiss}, \citenamefont {Xie} \emph {et~al.}}]{feder2025fluorescent}%
  \BibitemOpen
  \bibfield  {author} {\bibinfo {author} {\bibfnamefont {J.~S.}\ \bibnamefont {Feder}}, \bibinfo {author} {\bibfnamefont {B.~S.}\ \bibnamefont {Soloway}}, \bibinfo {author} {\bibfnamefont {S.}~\bibnamefont {Verma}}, \bibinfo {author} {\bibfnamefont {Z.~Z.}\ \bibnamefont {Geng}}, \bibinfo {author} {\bibfnamefont {S.}~\bibnamefont {Wang}}, \bibinfo {author} {\bibfnamefont {B.~B.}\ \bibnamefont {Kifle}}, \bibinfo {author} {\bibfnamefont {E.~G.}\ \bibnamefont {Riendeau}}, \bibinfo {author} {\bibfnamefont {Y.}~\bibnamefont {Tsaturyan}}, \bibinfo {author} {\bibfnamefont {L.~R.}\ \bibnamefont {Weiss}}, \bibinfo {author} {\bibfnamefont {M.}~\bibnamefont {Xie}}, \emph {et~al.},\ }\bibfield  {title} {\bibinfo {title} {A fluorescent-protein spin qubit},\ }\href@noop {} {\bibfield  {journal} {\bibinfo  {journal} {Nature}\ ,\ \bibinfo {pages} {1}} (\bibinfo {year} {2025})}\BibitemShut {NoStop}%
\bibitem [{\citenamefont {Meng}\ \emph {et~al.}(2025)\citenamefont {Meng}, \citenamefont {Nie}, \citenamefont {Berger}, \citenamefont {Nick}, \citenamefont {Einholz}, \citenamefont {Rizzato}, \citenamefont {Schleicher},\ and\ \citenamefont {Bucher}}]{meng_nie_berger_nick_einholz_rizzato_schleicher_bucher_2025}%
  \BibitemOpen
  \bibfield  {author} {\bibinfo {author} {\bibfnamefont {K.}~\bibnamefont {Meng}}, \bibinfo {author} {\bibfnamefont {L.}~\bibnamefont {Nie}}, \bibinfo {author} {\bibfnamefont {J.}~\bibnamefont {Berger}}, \bibinfo {author} {\bibnamefont {Nick}}, \bibinfo {author} {\bibfnamefont {C.}~\bibnamefont {Einholz}}, \bibinfo {author} {\bibfnamefont {R.}~\bibnamefont {Rizzato}}, \bibinfo {author} {\bibfnamefont {E.}~\bibnamefont {Schleicher}},\ and\ \bibinfo {author} {\bibfnamefont {D.~B.}\ \bibnamefont {Bucher}},\ }\bibfield  {title} {\bibinfo {title} {Optically detected and radio wave-controlled spin chemistry in cryptochrome},\ }\bibfield  {journal} {\bibinfo  {journal} {bioRxiv (Cold Spring Harbor Laboratory)}\ }\href {https://doi.org/https://doi.org/10.1101/2025.04.16.649006} {https://doi.org/10.1101/2025.04.16.649006} (\bibinfo {year} {2025})\BibitemShut {NoStop}%
\bibitem [{\citenamefont {Mann}\ \emph {et~al.}(2025)\citenamefont {Mann}, \citenamefont {Cowley-Semple}, \citenamefont {Bryan}, \citenamefont {Huang}, \citenamefont {Heutz}, \citenamefont {Attwood},\ and\ \citenamefont {Bayliss}}]{mann_chemically_2025}%
  \BibitemOpen
  \bibfield  {author} {\bibinfo {author} {\bibfnamefont {S.~K.}\ \bibnamefont {Mann}}, \bibinfo {author} {\bibfnamefont {A.}~\bibnamefont {Cowley-Semple}}, \bibinfo {author} {\bibfnamefont {E.}~\bibnamefont {Bryan}}, \bibinfo {author} {\bibfnamefont {Z.}~\bibnamefont {Huang}}, \bibinfo {author} {\bibfnamefont {S.}~\bibnamefont {Heutz}}, \bibinfo {author} {\bibfnamefont {M.}~\bibnamefont {Attwood}},\ and\ \bibinfo {author} {\bibfnamefont {S.~L.}\ \bibnamefont {Bayliss}},\ }\bibfield  {title} {\bibinfo {title} {Chemically {Tuning} {Room} {Temperature} {Pulsed} {Optically} {Detected} {Magnetic} {Resonance}},\ }\href {https://doi.org/10.1021/jacs.5c05505} {\bibfield  {journal} {\bibinfo  {journal} {Journal of the American Chemical Society}\ }\textbf {\bibinfo {volume} {147}},\ \bibinfo {pages} {22911} (\bibinfo {year} {2025})},\ \bibinfo {note} {publisher: American Chemical Society}\BibitemShut {NoStop}%
\end{thebibliography}%

\end{document}